\RequirePackage[undo-recent-deprecations]{expl3}
\documentclass[a4paper,fleqn,5p,times]{cas-dc}
\usepackage[numbers]{natbib}
\usepackage{amsmath}
\usepackage{algorithm}
\usepackage{algorithmic,amsmath}
\usepackage{float,graphicx} 
\usepackage{amsmath,amsfonts}
\usepackage{algorithmic}
\usepackage{graphicx}
\usepackage{subcaption} 
\usepackage{textcomp,lineno,hyperref,chngcntr,caption,wasysym,textcomp,mathrsfs,indentfirst,verbatim,url} 
\usepackage{verbatim}
\usepackage{algorithm}
\usepackage{algorithmic}
\usepackage{diagbox}
\usepackage{multirow}
\usepackage{threeparttable}
\newcommand{\tabincell}[2]{\begin{tabular}{@{}#1@{}}#2\end{tabular}} 

\def\tsc#1{\csdef{#1}{\textsc{\lowercase{#1}}\xspace}}
\tsc{WGM}
\tsc{QE}
\tsc{EP}
\tsc{PMS}
\tsc{BEC}
\tsc{DE}

\begin{document}
\let\WriteBookmarks\relax
\def\floatpagepagefraction{1}
\def\textpagefraction{.001}
\shorttitle{Improving Utility of Differentially Private Mechanisms}
\shortauthors{Wen et al. }
\let\printorcid\relax

\title [mode = title]{ Improving Utility of Differentially Private Mechanisms through Cryptography-based Technologies: a Survey}    
\author[1]{Wen Huang}
\author[1]{Shijie Zhou}
\author[2]{Tianqing Zhu} 
\author[1]{Yongjian Liao}
\address[1]{University of Electronic Science and Technology of China, Chengdu, Sichuan, P.R.China}
\address[2]{University of Technology Sydney,Sydney, Australia}

\begin{abstract}
Due to successful applications of data analysis technologies in many fields, various institutions have accumulated a large amount of data to improve their services. As the speed of data collection has increased dramatically over the last few years, an increasing number of users are growing concerned about their personal information. Therefore, privacy preservation has become an urgent problem to be solved. Differential privacy as a strong privacy preservation tool has attracted significant attention. In this survey, we focus on improving utility of between differentially private mechanisms through technologies related to cryptography. In particular, we firstly focus on how to improve utility through anonymous communication. Then, we summarize how to improve utility by combining differentially private mechanisms with homomorphic encryption schemes. Next, we summarize hardness results of what is impossible to achieve for differentially private mechanisms' utility from the view of cryptography. Differential privacy borrowed intuitions from cryptography and still benefits from the progress of cryptography. To summarize the state-of-the-art and to benefit future researches, we are motivated to provide this survey.
\end{abstract}


\maketitle

\section{Introduction}
\par Due to successful applications of data analysis techniques such as machining learning and deep learning, various institutions such as bank, hospital, and commercial companies start to accumulate data to improve their services including the field of climate \cite{198_overpeck2011climate}, biology \cite{199_marx2013biology}, health \cite{200_lang2011advancing}, and social science \cite{201_king2011ensuring}.However, these massive datasets always have sensitive information on individuals, especially these databases accumulated by companies that serve our daily life. Privacy preservation has become an urgent problem that needs to be solved and has drawn significant attention all over the world. For example, all major countries in the world enact laws to protect privacy of their citizens such as Data Protection Directive of European Union \cite{202_robinson2009review}, General Data Protection Regulation of European Union \cite{203}, California Consumer Privacy Act of American \cite{204} and  Internet Security Law of China \cite{205_deng2017consultative}. 

\par Differential privacy is proposed by Dwork et al. via a sequence of papers \cite{210_dinur2003revealing} \cite{211_dwork2004privacy} \cite{212_blum2005practical}. It becomes a widely accepted standard of privacy preservation nowadays for its promising capability to preserve privacy of scale data \cite{206_cao2018quantifying} and a strong privacy guarantee. For example, many big organizations have applied differential privacy to their commercial products including Google \cite{207_erlingsson2014rappor}, Uber \cite{208_johnson2018towards}, US Census Bureau \cite{209_machanavajjhala2008privacy}, and so on. 

\par Differential privacy benefits from cryptography \cite{0_dwork2006differential}. Specifically, Dwork proves the impossibility of intuition that "access to a statistical database should not enable one to learn anything about an individual that could not be learned without the access". The intuition is called semantic security defined by Goldwasser and Micali for cryptosystems \cite{213_goldwasser1984probabilistic}. By relaxing the concept of indistinguishability of cryptography to $\epsilon$-indistinguishability, Dwork introduces differential privacy. In particular, differential privacy does not require that attackers cannot distinguish two databases absolutely, but require that attackers distinguish two databases with an extremely small probability such that security of differential privacy could be accepted by users. 

\par Various upcoming applications are asking differentially private mechanisms for higher data utility. For example, mechanisms of centralized differential privacy add prohibitively high noise to final results because of global sensitivity. Then, to reduce added noise, many new concepts of sensitivity such as smooth sensitivity \cite{189_nissim2007smooth}, elastic sensitivity \cite{208_johnson2018towards} are proposed. To eliminate the trusted center, the centralized differential privacy shifts to local differential privacy. However, to preserve user privacy, mechanisms of local differential privacy add so much noise to databases, resulting in low data utility \cite{234_murakami2019utility}. In addition, more complex data analysis tasks such as machine learning \cite{235_ji2014differential} and deep learning \cite{236_abadi2016deep} encounter privacy problem. These tasks are demanding higher data utility from differentially private mechanisms such that differentially private mechanisms can be applied to solve privacy problems in these tasks.    

\par Some technologies related to cryptography are promising to improve utility of differentially private mechanisms by recent researches such as anonymous communication technology \cite{37_cheu2019distributed}  \cite{36_bittau2017prochlo} \cite{35_erlingsson2019amplification} and homomorphic encryption schemes \cite{46_kim2019secure} \cite{181_kacsmardifferentially} \cite{186_acar2017achieving}. Therefore, we are motivated to provide this survey to summarize how to improve utility of differential privacy and to benefit future researches about improving utility of differential privacy by cryptography.

\subsection{Other Surveys}
\par Differential privacy has developed for many years and has been applied to various fields. So far, there have been multiple survey papers focusing on different aspects of differential privacy. 
\par \textbf{Surveys on developments of differential privacy}

\par $\bullet$ Dwork recalls the definition of differential privacy and two basic techniques for achieving it \cite{237_dwork2008differential}.
\par $\bullet$ Dwork summarizes prospective solutions for difficulties which arise from the context of statistics analysis \cite{238_dwork2010differential}.
\par $\bullet$ Dwork summarizes motivating scenarios and future directions of differential privacy \cite{239_dwork2011firm}. 
\par $\bullet$ A book of Dwork and Roth presents fundamental algorithms and mainly theories which are widely used in differential privacy. The book is a good starting point for beginners \cite{220_dwork2014algorithmic}.
\par $\bullet$ Vadhan gives a tutorial about the complexity of differential privacy \cite{42_vadhan2017complexity}.
\par $\bullet$ Desfontaines makes a comprehensive survey of differential privacy in the lens of seven dimensions \cite{217_desfontaines2020sok}.
\par \textbf{Surveys on applications of differential privacy}
\par $\bullet$ Yu talks about the differential privacy in the lens of big data \cite{219_yu2016big}.
\par $\bullet$ Zhu et al. make a survey about applications of differential privacy in data publishing and analyses\cite{218_zhu2017differentially}.
\par $\bullet$ Yang et al. survey local differential privacy in the field of statistical query and private learning \cite{214_yang2020local}.
\par $\bullet$ Zhao et al. survey potential applications of local differential privacy to the internet of connected vehicles \cite{215_zhao2019survey}.

\par $\bullet$ Hassan et al. investigate applications of differential privacy to Cyber Physical Systems \cite{216_hassan2019differential}.

\begin{figure*}[pos=htbp]
	\centering
	\includegraphics[scale=0.29]{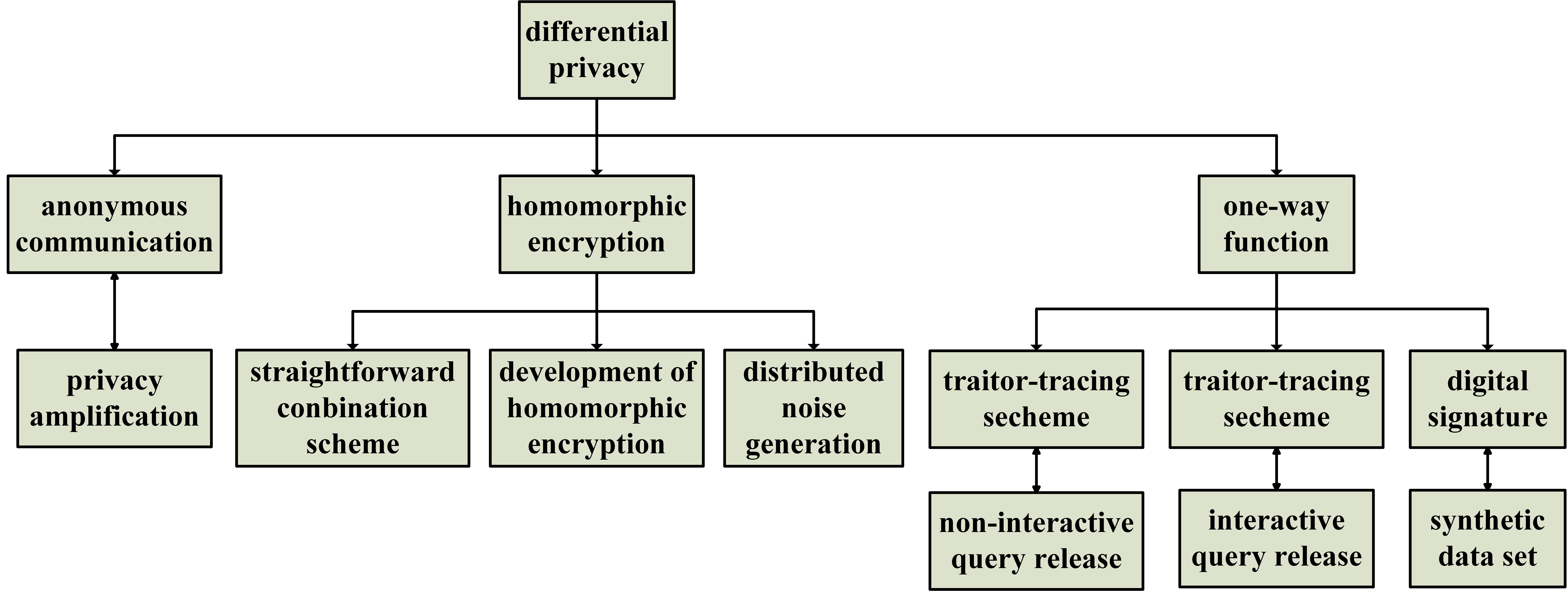}
	\subcaption*{Fig 1: Survey Structure }	 
\end{figure*}

\par To the best of our knowledge, this is the first survey that focuses on how to improve utility through cryptography. We hope that this survey can benefit future researches about improving utility of differentially private mechanisms.

\subsection{Organization}
\par In this survey, we focus on the connection between differentially private mechanisms and three technologies including anonymous communication, homomorphic encryption and one-way function. In particular, the anonymous communication technology and homomorphic encryption schemes can improve utility of differentially private mechanisms. The one-way function is the foundation of proving what is impossible to achieve for differential privacy's utility by standard cryptographic assumptions. The structure of this survey is shown in Fig 1.  


\par In the next section, preliminaries of differential privacy are introduced. Then, in the third section, utility amplification through anonymous communication is presented, and connections among differential privacy, anonymous communication, and $k$-anonymity are discussed. In the fourth section, utility amplification through homomorphic encryption schemes is presented. In the fifth section, hardness results of utility are demonstrated with aspect to non-interactive query release, interactive query release, and synthetic databases.  In the sixth section, we share out future directions. At last, we conclude this survey.

\section{Preliminaries}
\par Differential privacy is a strong privacy concept. It formulates the intuition that presence of one's data record in a database should not increase the risk of leaking his private information. To that end, differentially private mechanisms guarantee that its outputs are insensitive to any particular data record. Two databases differing on one record are called neighboring databases denoted by $D \sim D'$.     

\par \textbf{Definition 1} (Differential privacy) Any random mechanism $M:W^n\to R^d$ preserves $(\epsilon,\delta)$-DP(differential privacy) if for any neighboring databases $D$ and $D'$ as well as for all sets of possible outputs $S$:
\begin{eqnarray*}
	P\{M(D)\in S\} \le e^\epsilon P\{M(D')\in S\} + \delta
\end{eqnarray*}
\par Here, $\epsilon$ is the privacy budget. The privacy guarantee is quantified by the privacy budget. The smaller the privacy budget is, the stronger the privacy guarantee is. $\delta$ is an extremely small number which is cryptographically negligible. It can be interpreted as an upper-bound on the probability of catastrophic failure.

\subsection{Basic Models}

\par According to whether there is a trusted center, differential privacy models are divided into two types, namely centralized differential privacy model and local differential privacy model. In centralized differential privacy model, there is a trusted aggregator who collects data from users and perturbs results with noise before final results are distributed. On the contrary, in local differential privacy model, there is not a trusted aggregator. Each participant perturbs his data before the data are sent to the aggregator such that each participant's privacy can be preserved even if the aggregator possibly attempts to infer information about participants. The difference between centralized differential privacy model and local differential privacy model is shown in Fig 2.   

\begin{figure*}[pos=htbp]
	\centering	
	
	\begin{minipage}[t]{0.48\linewidth}
		\centering
		\includegraphics[width=6.5cm]{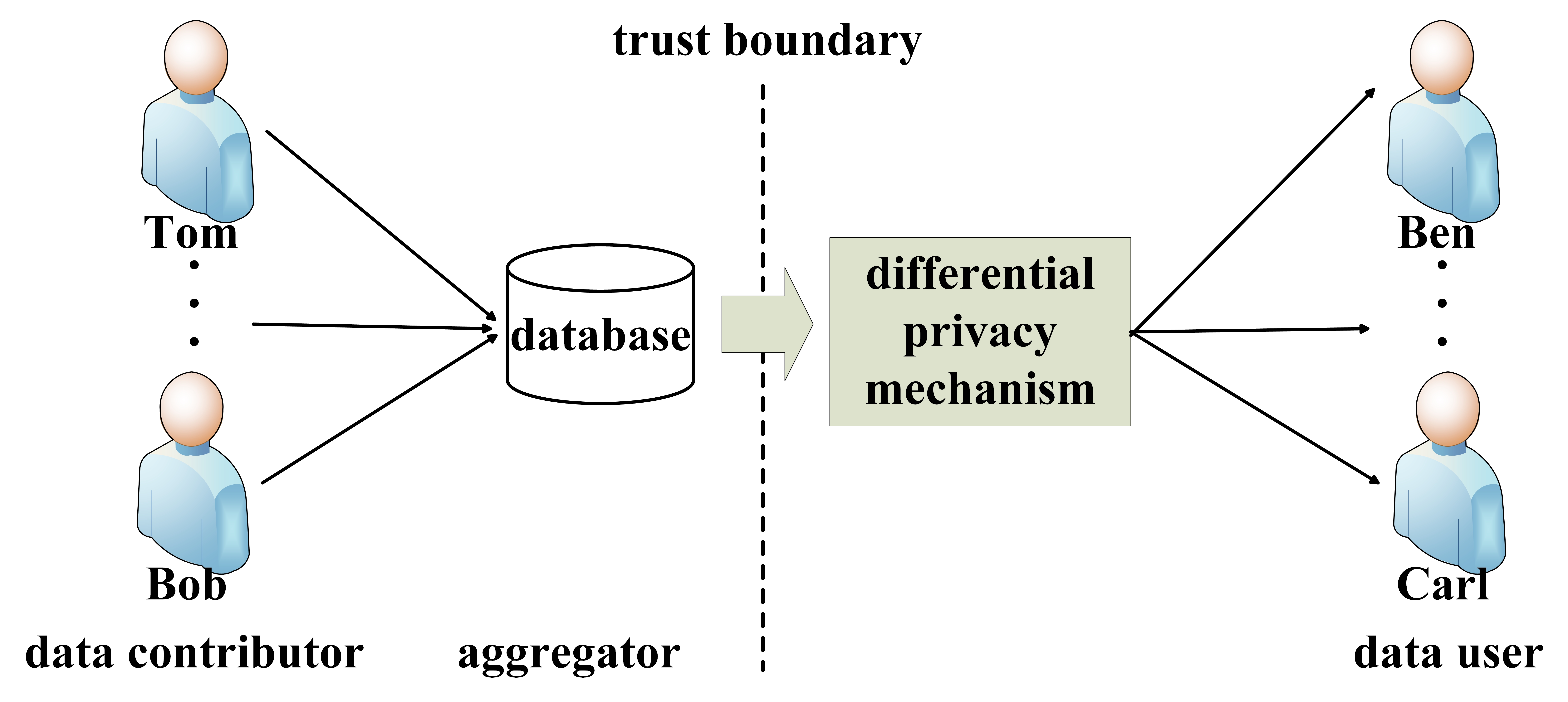}\\
		\subcaption*{(a) centralized differential privacy model}			 	 
	\end{minipage} 	
	\begin{minipage}[t]{0.48\linewidth}
		\centering
		\includegraphics[width=6.5cm]{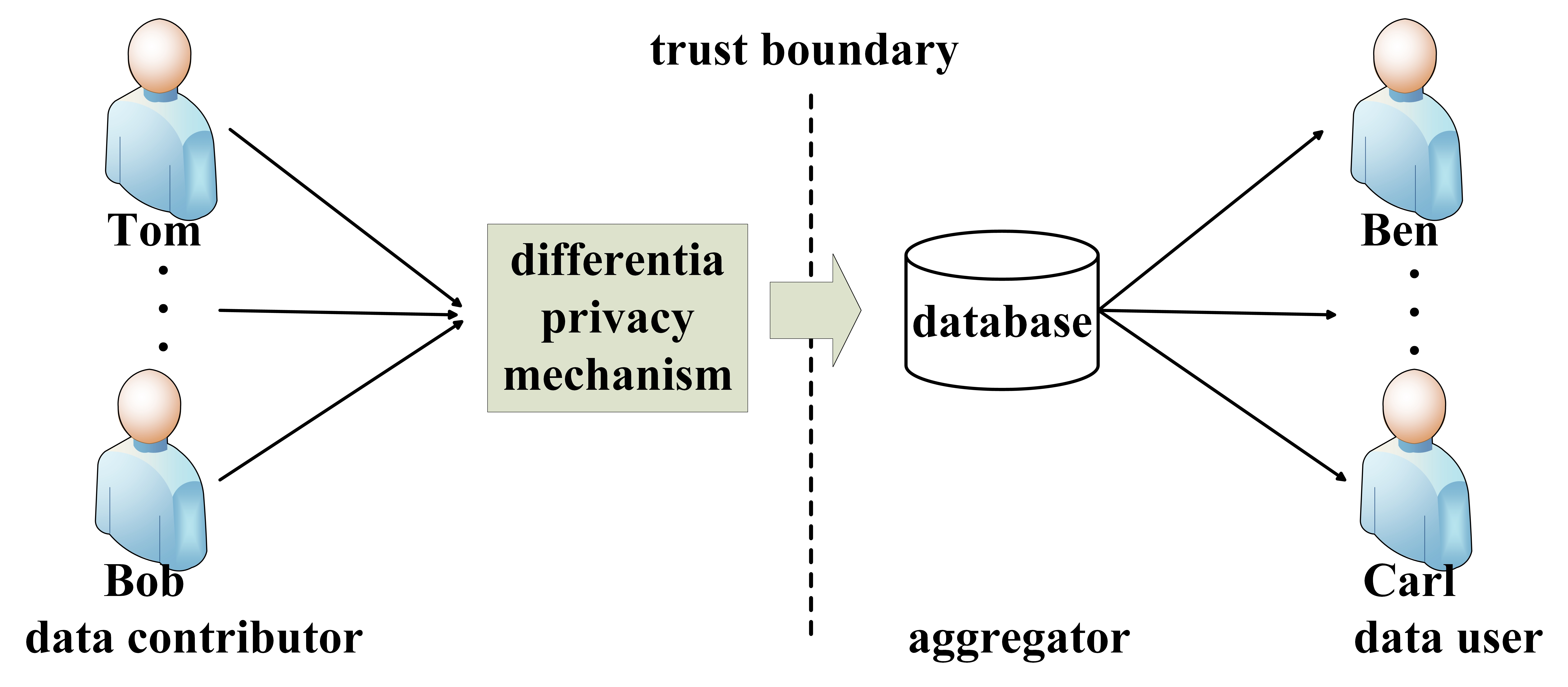}\\
		\subcaption*{(b) local differential privacy model }		 		 
	\end{minipage}	
	
	\centering
	\subcaption*{Fig 2: Comparison of Two Differential Privacy Models}
	
	\label{fig:compare_fig}	
	
\end{figure*}

\subsection{Basic Mechanisms}

\par To realize privacy intuitions of differential privacy, differentially private mechanisms perturb their output with noise such that the difference caused by the presence or absence of any data records can be masked by added noise. Therefore, a concept named global sensitivity is introduced to quantify the biggest difference on the final output, which is caused by the presence or absence of one record. Formally,

\par \textbf{Definition 2} (Global sensitivity \cite{224_dwork2011firm}) For a given database $D$ and query $q$, the global sensitivity denoted by $\Delta D$ is	
\begin{eqnarray*}
	\Delta D = \max \limits_{D \sim D'} |q(D)-q(D')|
\end{eqnarray*}  

\par Although any mechanisms satisfying Definition 1 could be regarded as differentially private mechanisms, there are four widely used differentially private mechanisms, including the Laplace mechanism, the Gaussian mechanism, exponential mechanism, and randomized response mechanism. 

\par The Laplace mechanism and the Gaussian mechanism are to deal with numerical data. Specifically, the Laplace mechanism perturbs its output with noise from the Laplace distribution with the location parameter of 0 and the scale parameter of $\frac{\Delta D}{\epsilon}$. Formally,    
\par \textbf{Definition 3} (Laplace mechanism \cite{223_dwork2016calibrating}) For a given database $D$, a query $q$, and a privacy budget $\epsilon$, the output of the Laplace mechanism is 
\begin{eqnarray*}
	M(q,D,\epsilon) = q(D) + Lap(0,\frac{\Delta D}{\epsilon}) 
\end{eqnarray*} 
\par Here, $Lap(0,\frac{\Delta D}{\epsilon})$ represents added noise from the Laplace distribution with the location parameter of 0 and the scale parameter of $\frac{\Delta D}{\epsilon}$. Notably, $\delta$ is 0 for the Laplace mechanism.

\par The Gaussian mechanism covers the difference on queries' answer over neighboring databases by noise from the Normal distribution with the location parameter $\mu=0$ and the scale parameter $\sigma=\sqrt{2(\ln1.25-\ln\delta)}\Delta D/\epsilon$. Formally,  

\par \textbf{Definition 4} (Gaussian mechanism \cite{220_dwork2014algorithmic}) For a given data set $D$, a query $q$ and a privacy budget $\epsilon$, the output of the Gaussian mechanism is 
\begin{eqnarray*}
	M(q,D,\epsilon,\delta) = q(D) + N(0, \sigma) 
\end{eqnarray*} 
\par Here, $N(0,\sigma)$ represents added noise from the Normal distribution with the location parameter of 0 and the scale parameter of $\sigma$. 

\par As for the exponential mechanism, it is based on a function called the utility function which is to measure quality of outputs. Inputs of the utility function are a database $D$ and a possible output $r$. The utility function should be insensitive to the present or absence of any one record. The sensitivity of an utility function $u$ is $\Delta u$
\begin{eqnarray*}
	\Delta u = \max \limits_{\forall r,D, D'} |u(D,r)-u(D',r)|  
\end{eqnarray*} 

\par The exponential mechanism is for data which become meaningless when these data are perturbed by numerical noise such as class attributions, say male. The exponential mechanism assigns the greatest probability to a result whose value of the utility function is the greatest. Formally,  

\par \textbf{Definition 5} (Exponential mechanism \cite{222_mcsherry2007mechanism}) For an utility function $u$ and a database $D$, the output of the exponential mechanism is 
\begin{eqnarray*}
	M(D,q) = \{return\ r\ with\ probability \propto e^{\frac{\epsilon q(D,r)}{2\Delta u}}\} 
\end{eqnarray*} 

\par The randomized response is suitable for applications with multiple users, which is proposed by Warner et al. as a survey technology to eliminate the evasive answer bias \cite{221_warner1965randomized}. Formally, 

\par \textbf{Definition 6} (Generalized randomized response \cite{214_yang2020local}) Given a user with a value $v \in R$, where R is a set of $d$ possible true values that the user can have. A random variable, denoted by $\hat{t}$, represents the response of a user with sample space $R$: The generalized randomized response works as follow: 
\begin{eqnarray*}
	P\{\hat{t} = v\} = 
	\left\{  
	\begin{aligned}
		\frac{e^\epsilon}{e^\epsilon + d -1}\ if\ t=v\\
		\frac{1}{e^\epsilon + d -1}\ if\ t \ne v
	\end{aligned}
	\right.  
\end{eqnarray*}

\section{Improving Utility through Anonymous Communication}

\par The anonymous communication was proposed by Chaum et al. for electronic mail \cite{100_chaum1981untraceable}. E-mails from one person to another may be with some unique identifiers such as a source IP address or a MAC address of Ethernet. These identifiers are available to any one who observes the communication channels of E-mails. Therefore, the observer may link transactions of E-mails to a certain address, resulting in that transactions of E-mails could be linked to a certain person. However, the linkage between a person and E-mails could be a big privacy breach. The anonymous communication aims for making sure that transactions over data channels cannot be linked to a subject such as one device or one person \cite{101_shirazi2018survey}. 

\par Anonymity is the privacy foundation of anonymous communication. The formal definition of anonymity is: the state of being not identifiable within a set of subjects \cite{102_danezis2008survey}. As can be seen from the definition, the concept of anonymity attempts to capture the privacy intuition of "hiding in the crowd". The intuition is common in the filed of privacy preservation. For example, the concept of $k$-anonymity also attempts to capture the intuition \cite{103_sweeney2002k}. 
\begin{figure}[pos=htbp]
	\centering
	\includegraphics[scale=0.3]{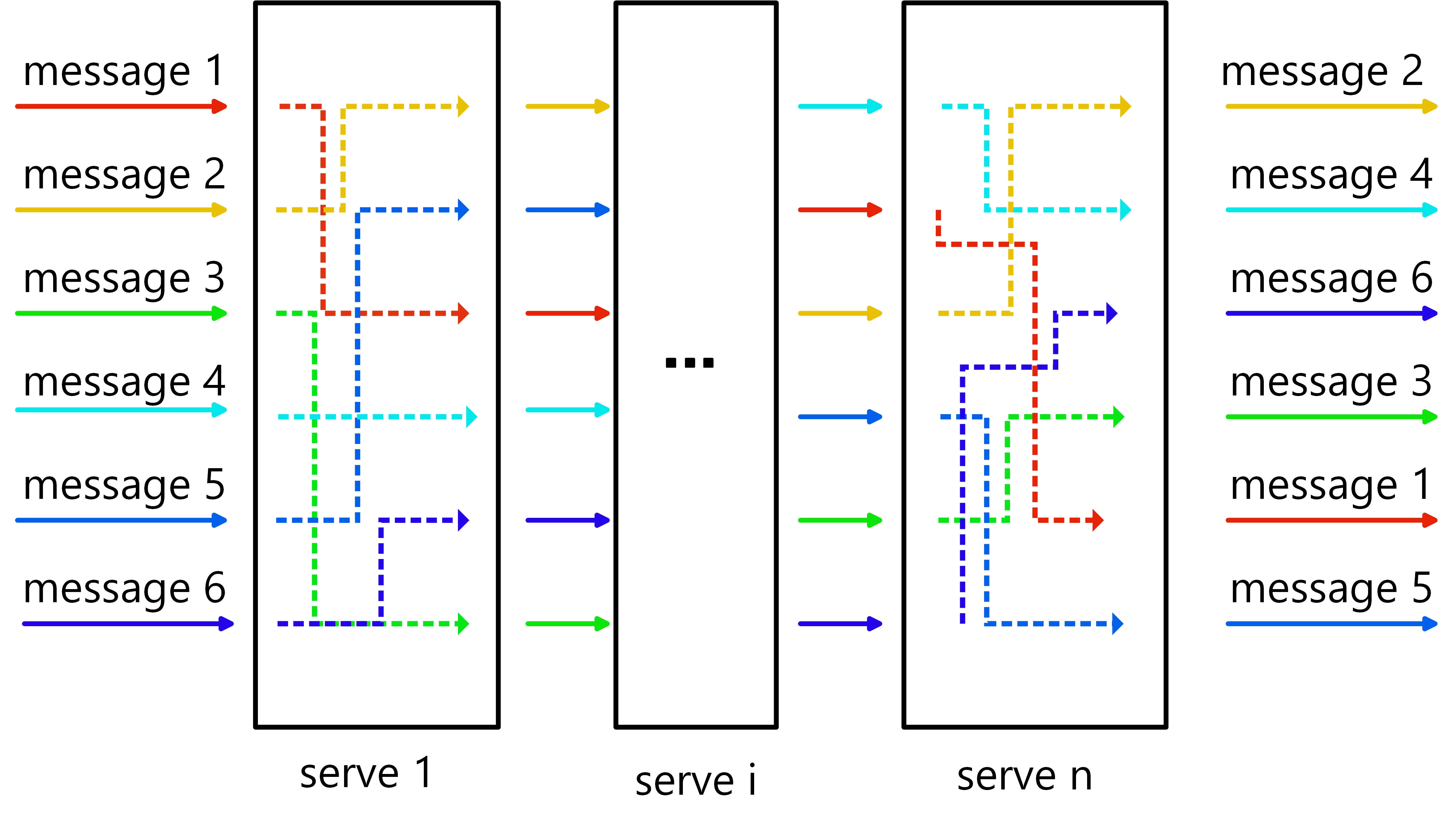}
	\subcaption*{Fig 3: Mix Networks}	
\end{figure} 

\par There are some approaches to achieve anonymous communication, for instance, Mix Networks in Fig 3 \cite{100_chaum1981untraceable}. Mix Networks consist of a sequence of servers whose encryption keys and order are public information. Messages to be transmitted are encrypted by encryption keys of all servers one by one in order from $n$ to 1. When encrypted messages are submitted to the first server, it decrypts all these messages and permutes these decrypted messages randomly. Then, the first server sends all decrypted messages to the next server. Every server performs the same operations until the last server sends messages to the target address.      

\subsection{Utility Amplification}         

\par Local differential privacy suffers from big noise added by all participants so many papers focus on improving its utility such as \cite{105_gursoy2019secure} \cite{104_alvim2018local}. In particular, every participant of local differential privacy adds random noise to data held by himself so that privacy of participants can be preserved. This approach results in big noise in total, for instance $O(\sqrt{N})$ for the case of statistical counting queries \cite{37_cheu2019distributed}. Papers such as \cite{37_cheu2019distributed}  \cite{36_bittau2017prochlo} \cite{35_erlingsson2019amplification} find that combining local differential privacy with anonymous communication can amplify the utility guarantee of local differential privacy.   

\par Bittau et al. propose a system architecture ---Encode, Shuffle, Analyze(ESA)--- for performing large-scale monitoring such as error reporting of software and demographic profiling with high utility while also protecting user privacy \cite{36_bittau2017prochlo}. The system architecture is shown in Fig 4. 
\begin{figure}[pos=htbp]
	\centering
	\includegraphics[scale=0.35]{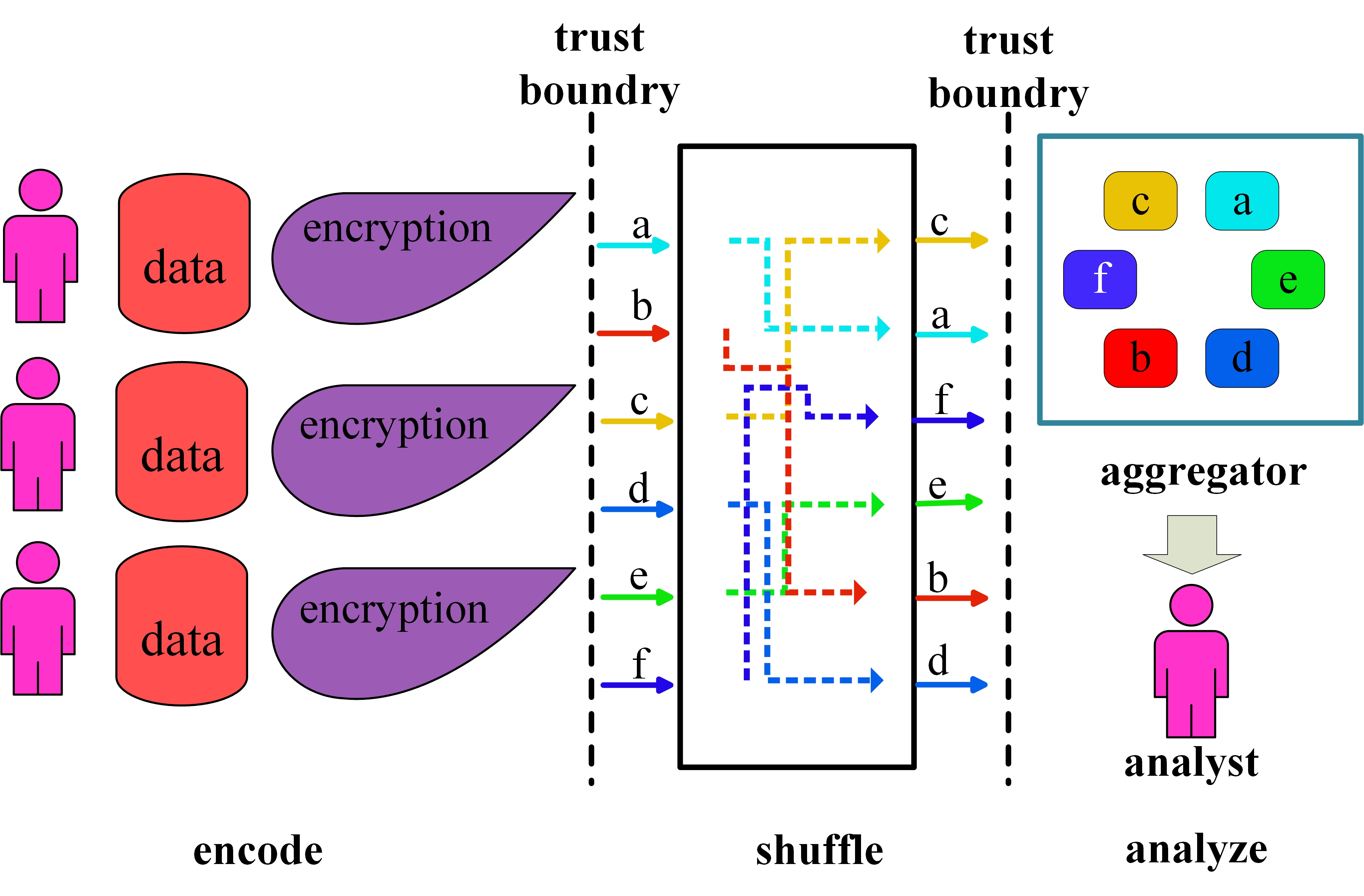}
	\subcaption*{Fig 4: ESA Architecture: Encode, Shuffle, and Analyze.}	
\end{figure} 
\par The pipeline of ESA consists of three steps logically. In the step of encoding, data are encoded. The encoding scheme depends on application scenarios. For example, the encoding scheme could be the binary code when the task is to obtain the frequency of strings like \cite{106_fanti2016building}. The encoding scheme also could be raw data with noise calibrated by requirements of local differential privacy if the target is to preserve privacy of users by local differential privacy like \cite{107_wang2019collecting}. In the step of shuffling, trusted proxy takes all participants' messages within a time epoch as inputs and then outputs these messages in random order. Shuffling makes sure that each record hides in the crowd. This can be implemented by anonymous communication such as Mix Networks \cite{100_chaum1981untraceable} mentioned before. In the step of analyzing, records are collected and analyzed by various algorithms such as \cite{108_hayes2018contamination}.    

\par The system architecture ESA and its implement Prochlo are a flexible framework. They can be used to perform multiple data analysis tasks while preserving privacy of data owners. Bittau et al. find that shuffling can amplify the utility guarantee of local differential privacy, and they demonstrate the phenomenon by extensive experiments \cite{36_bittau2017prochlo}.   

\par Erlingsson et al. improve the utility guarantee of local differential privacy by shuffling for the frequency estimation problem \cite{35_erlingsson2019amplification}. Specifically, all participants in the estimation task send one bit of 0 or 1 to an aggregator who compute the sum of received bits as the frequency. By anonymous communication which is used to randomly shuffle all bits of participants, the privacy cost can be reduced by a scale about $O(\sqrt{n})$. $n$ is the number of participants who obey the proposed mechanism. That is, $n$ is the number of participants who is semi-honest(honest but curious) in term of security assumptions. In particular, for a $\epsilon$ local differential privacy mechanism, if data of participants are shuttled randomly by anonymous communication before data are sent to the aggregator, the mechanism could become a $(e^\epsilon -1)e^{2\epsilon}\sqrt{log(1/\delta)/n}$ central differentially private mechanism.     

\par Cheu et al. also propose a mechanism for the frequency estimation problem \cite{37_cheu2019distributed}. Their positive results are similar to \cite{35_erlingsson2019amplification}. Cheu et al. also extend their mechanism to estimate a bounded real-valued statistic with the cost of additional communication overhead. They also analyze shuffling in applications of histograms and variable-selection. By these analyses, they empirically conclude that the central, shuffled, and local models are strictly ordered in the accuracy they can achieve. The proposed mechanism is based on Prochlo \cite{36_bittau2017prochlo} and shuffling is implemented by anonymous communication technology.  

\par Wagh et al. \cite{2_wagh2020dp} present a way to understand utility amplification of anonymous communication technology from the sub-sampling argument \cite{109_balle2018privacy}  \cite{110_kasiviswanathan2011can}.  The improved privacy guarantee can be demonstrated by a simulation algorithm with two steps. In the first step, a value is sampled from a binomial distribution $B(N,p)$, which is the number of participants who send a random bit of 0 or 1 to a aggregator. In the second step, a subset of all received bits is sampled, and the final result is computed on the chosen subset. Uncertainty introduced by the sampling process amplifies the privacy guarantee. 

\subsection{Discussions}	 

\par Anonymous communication formulates the privacy intuition of  "hiding in the crowd" in terms of communication processes. Another privacy concept called $k$-anonymity also formulates the privacy intuition of "hiding in the crowd" although it is for contents of data records \cite{103_sweeney2002k}.  

\subsubsection{Improving Utility through $k$-anonymity}
\par $k$-anonymity aims at making every record indistinguishable from at least other $k-1$ records in a dataset by manipulating attribution values of data records. Attributions of a data record are divided into two classes in $k$-anonymity mechanisms. The class of attributions which could be used to identify the owner of records is named quasi-identifiers. Every $k$-anonymity mechanism attempts to suppress or generalize these quasi-identifiers so that each record is indistinguishable from at least other $k-1$ records. For example, Fung et al. build their method on generalization (replacing data values with more general values) \cite{111_fung2005top} and Kisilevich et al. build their method on suppression(not releasing a value at all) \cite{112_kisilevich2009efficient}.  

\par The concept of $k$-anonymity has generally be considered too weak. In particular, while $k$-anonymity prevents disclosure of identity information, it is insufficient to prevent the disclosure of attributes \cite{113_li2007t}. Thus, the other privacy concepts are introduced such as $l$-diversity \cite{114_machanavajjhala2007diversity} and $t$-closeness \cite{113_li2007t}. However, all these privacy concepts including $k$-anonymity, $l$-diversity, and $t$-closeness suffer from background knowledge of attackers. When attackers have appropriate background knowledge, these privacy concepts always can be broken \cite{115_domingo2015t}. Differential privacy is introduced as the privacy concept which is independent of attackers' background knowledge. Independence of attackers' background knowledge is one of important reasons for popularity of differential privacy.

\par Li et al. find that $k$-anonymity can amplify the utility guarantee of differential privacy\cite{38_li2011provably}. In particular, they assume that there is a "safe" $k$-anonymous scheme which can guarantee that every record is indistinguishable from at least other $k-1$ records no matter what kind of background knowledge attackers possess. Their mechanism is a three steps pipeline. First, raw data records are randomly sampled with a certain probability. Then, sampled data records are processed by the safe $k$-anonymity scheme. At last, these anonymous data records are processed by differentially private mechanisms. Li et al. show that combination between $k$-anonymity mechanisms and differentially private mechanisms can amplify the utility guarantee of differentially private mechanisms theoretically.        

\par Li's work \cite{38_li2011provably}, Erlingsson's work \cite{35_erlingsson2019amplification}, and Cheu's work \cite{37_cheu2019distributed} mutually provide support for each other's positive results. The reason is that the $k$-anonymity technology and anonymous communication technology actually formulate the same privacy intuition of "hiding in the crowd". The difference between these two technologies is that anonymity is for different objects. Specifically, for $k$-anonymity technology, anonymity is for objects who look at contents of data records. However, for anonymous communication technology, anonymity is for observers of communication channels. Although Li's work uses the subsampling technology which also can amplify the utility guarantee of differentially private mechanisms,  Erlingsson's work, and Cheu's work is also related to the subsampling technology from the lens of Wagh's work \cite{2_wagh2020dp}.
\subsubsection{Restrictions}
\par The application of anonymous communication to amplify the utility guarantee of differentially private mechanisms also has restriction. As mentioned in Li's work, the used $k$-anonymity scheme needs to be a "safe" $k$-anonymity scheme. That is, data records need to be really indistinguishable from at least other $k-1$ records. Similarly, used anonymous communication mechanisms are also required to guarantee that the intuition of "hiding in the crowd" can be realized.

\par However, this requirement for anonymous communication is hard to be satisfied when anonymous communication is combined with local differentially private mechanisms due to the basic difference between cryptography and differential privacy \cite{0_dwork2006differential} shown in Fig 5. In particular, there are three parties in cryptography including sender, receiver and attacker. Anonymous communication strips metadata of transactions such as the source IP address and the MAC address so that attackers who observe communication channels cannot determine whether transaction data are from a certain sender. But, in differential privacy's settings, there are only two parties including sender and receiver because receiver is also attacker who aims for inferring information of data owners. Thus, when anonymous communication is combined with differentially private mechanisms, it can stop attackers from inferring data owners by metadata such as the source IP address. However, anonymous communication cannot stop attackers from inferring data owners by contents of data records. Actually, this basic difference is one of motivations to promote the proposal of differential privacy in \cite{0_dwork2006differential}.        

\begin{figure}[pos=htbp]
	\centering
	\includegraphics[scale=0.4]{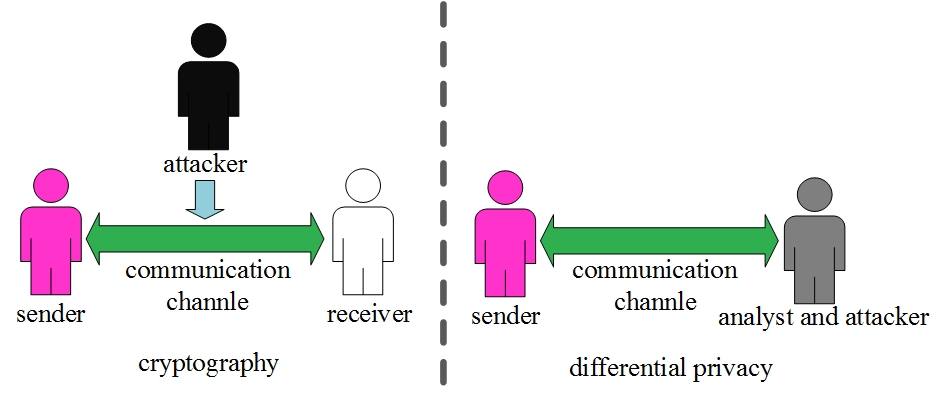}
	\subcaption*{Fig 5: Comparison between Cryptography and Differential Privacy}	
\end{figure} 

\par $k$-anonymity is the technology that formulates the same privacy intuition of anonymous communication technology in terms of contents of data records. However, $k$-anonymity fails to guarantee the privacy intuition, which is one of reasons accounting for popularity of differential privacy. Thus, directly applying the anonymous communication technology to make sure the privacy intuition of "hiding in the crowd"  may not work. In addition, these positive results in Erlingsson's work \cite{35_erlingsson2019amplification} and Cheu's work \cite{37_cheu2019distributed} hold due to special properties of their applications. Specifically, data records are bits of 0 or 1 in their setting. It is impossible to infer data owners by contents of data records. The anonymous communication technology strips metadata and randomly permutes order of data records, satisfying requirements of anonymous communication and requirements of $k$-anonymous at the same time.   

\par In a word, although there are some works which provide positive results about amplifying the utility guarantee of local differentially private mechanisms by anonymous communication, combination between anonymous communication and local differentially private mechanisms needs more further explorations. The anonymous communication only guarantees the implementation of its privacy intuition in terms of the communication layer. If attackers can infer information from contents of data records, the privacy guarantee of combining anonymous communication with local differentially private mechanisms may be compromised.  Although $k$-anonymity captures the same privacy intuition with anonymous communication, it suffers from background knowledge of attackers. Therefore, combining $k$-anonymity with local differentially private mechanisms may result in the possibility that the privacy guarantee of local differentially private mechanisms will be influenced by background knowledge of attackers.     

\section{Improving Utility through Homomorphic Encryption}

\par Homomorphic encryption is one kind of schemes in the field of secure multi-party computation. In particular, homomorphic encryption allows a third party to perform certain functions on encrypted data to obtain new encryption data such that these new obtained encryption data are the same with encryption data of plaintext obtained by performing identical functions on original plaintext \cite{43_acar2018survey}. Roughly speaking, the homomorphic encryption schemes allow exchanging execution order of encryption and certain functions. Formally, an encryption scheme is homomorphic over any function $f$ if the equality holds
\begin{eqnarray*}
	f(E(m_1),E(m_2)) = E(f(m_1,m_2)) \ \ \forall m_1,\ m_2 \in M
\end{eqnarray*}   
\par Where, $E$ is the encryption algorithm, and $M$ is the set of all possible messages. For example, if the function is addition, then the equality is $E(m_1) \diamondsuit E(m_2) =E(m_1+m_2)$. $\diamondsuit$ represents certain operations on encryption data and this new symbol is used to emphasize that operations on encryption data are not necessarily the same as operations on plaintext. 

\par Two motivations promote combination between  secure multi-party computation schemes and differentially private mechanisms. The first motivation is to improve utility of local differentially private mechanisms. In particular, each participant of local differentially private mechanisms perturbs his data locally before these data are sent to an aggregator such that aggregation results satisfy requirements of local differential privacy without dependence of a trust aggregator. However, due to added noise of each participant, utility of local differential privacy decreases dramatically, compared with centralized differentially private mechanisms. For example, the error boundary of local differentially private mechanisms is greater than the error boundary of centralized differentially private mechanisms with a multiplication factor $\sqrt{n}$ \cite{37_cheu2019distributed}. Secure multi-party computation schemes, such as homomorphic encryption, can simulate a trusted center such that the error boundary of local differentially private mechanisms can be reduced to the error boundary of centralized differentially private mechanisms.

\par The second motivation is to make final results of secure multi-party computation satisfying requirements of differential privacy so that attackers cannot infer individuals' information from these final results. Specifically, secure multi-party computation can make sure that any third parties and participants of computation can learn nothing except final results. For example, homomorphic encryption schemes enable executing operations on encryption data so that an aggregator can learn nothing except encryption results of final results. However, secure multi-party computation cannot stop information leakage from final results \cite{46_kim2019secure}. For example, two parties compute their average salaries via secure multi-party computation methods. When one party obtains their average salary, the salary of the other party can be inferred easily. The differentially private mechanisms can stop information leakage from final results so the combination between differentially private mechanisms and homomorphic encryption schemes can guarantee result privacy.          

\par In this section, we focus on the combination between differentially private mechanisms and homomorphic encryption schemes. Homomorphic encryption is more general than other secure multi-party computation schemes such as secrete sharing methods \cite{149_attasena2017secret} because homomorphic encryption can support more rich operations and functions such as the logarithm and exponential function. These functions are widely used in data analysis tasks such as the classification task \cite{150_huang2019efficient} in machine learning. In addition, data analysis tasks usually are based on many complex operations. Therefore, the combination between differentially private mechanisms and homomorphic encryption schemes are more nature and wide in real word applications.      

\subsection{Straightforward Schemes} 

\par There is a straightforward approach to combine differentially private mechanisms with homomorphic encryption schemes such that nothing can be learned except final results, and individuals' information cannot be inferred via these final results. The architecture of this straightforward scheme is shown in Fig 6.
\begin{figure}[pos=htbp]
	\centering
	\includegraphics[scale=0.35]{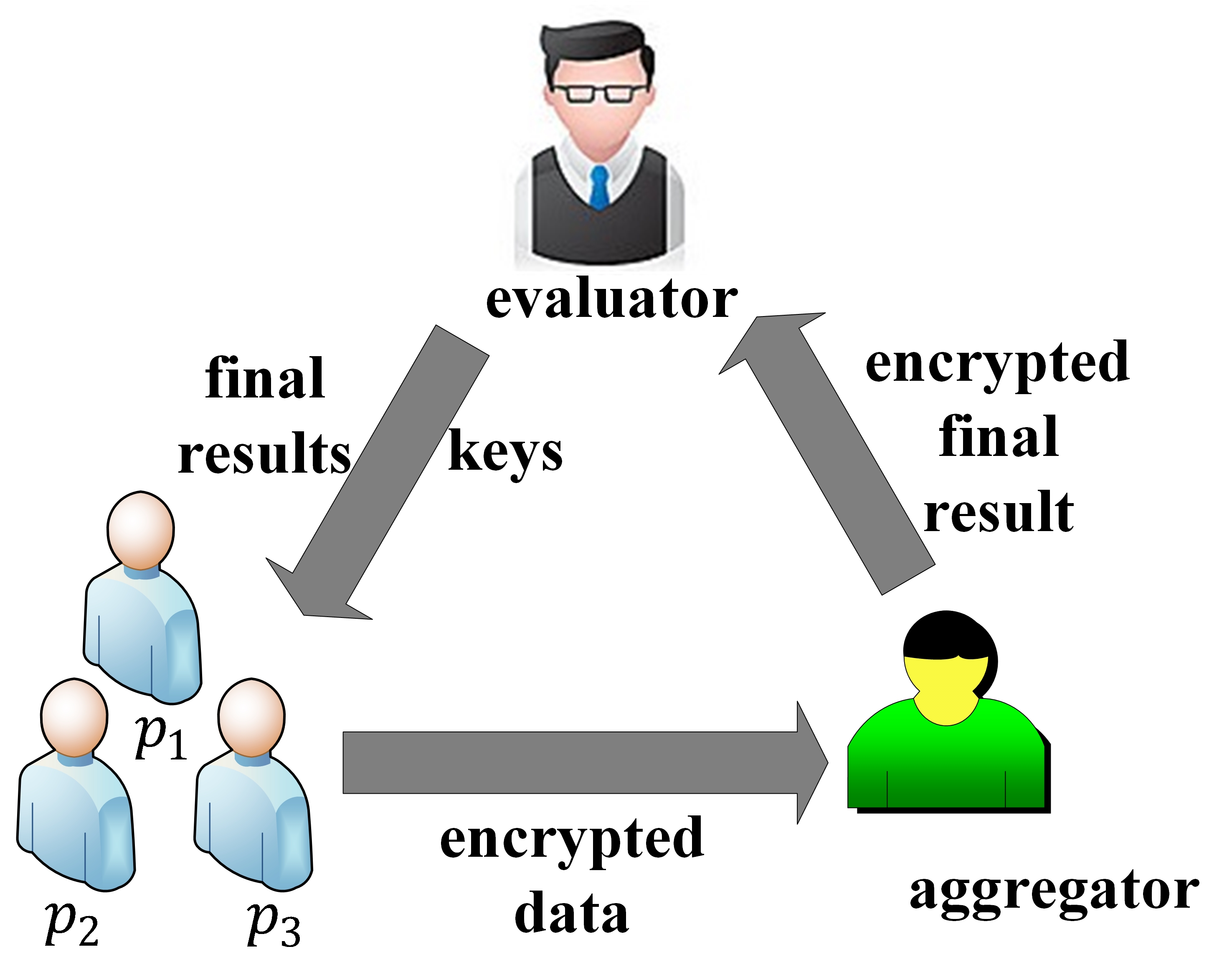}
	\subcaption*{Fig 6: Architecture of Straightforward Scheme}
\end{figure} 
\par $\bullet$ \textbf{Setup} The evaluator and each participant establish keys which they will use.
\par $\bullet$ \textbf{Update} Each participant encrypts his data with noise such that the summary of all participants' noise obeys the Laplace distribution(maybe other distributions, such as the Gaussian distribution, which can enable final results to satisfy requirements of differential privacy). These encrypted data are sent to an aggregator. 
\par $\bullet$ \textbf{Aggregation} The aggregator obtains each participants' encryption data and does some certain operations over all these encryption data to obtain the final result in encrypted form. Then, the encrypted final result is sent to the evaluator.
\par $\bullet$ \textbf{Evaluation} The evaluator decrypts these encryption data of the final result into plaitext and broadcasts the final result to each participant.  
\par There are two key components in the straightforward scheme, namely the distributed noise generation method and homomorphic encryption schemes. Because noise generation methods are tightly related to the error of final results, and chosen homomorphic encryption schemes are tightly related to the computation complexity and communication overhead.

\subsection{Homomorphic Encryption Schemes}

\par Homomorphic encryption has developed for decades. The term homomorphism was used firstly by Rivest et al. \cite{151_rivest1978data}. Their objective is to compute without decrypting encrypted data. After their work, there are numerous attempts to design homomorphic encryption schemes which can support rich operations and functions. Until 2009, Gentry has proposed the first fully homomorphic encryption scheme which can perform any computable functions on encryption data. \cite{152_gentry2009fully}.  

\par There are three types of homomorphic encryption schemes including partially homomorphic encryption schemes, somewhat homomorphic encryption schemes and fully homomorphic encryption schemes. Classification of homomorphic encryption schemes is based on supported operations including addition, multiplication, and the number of times operations can be performed. In particular, partially homomorphic encryption schemes allow performing one type of operation unlimited number of times such as addition  \cite{153_rivest1978method} \cite{154_10.1145/800070.802212} \cite{157_paillier1999public} or multiplication \cite{155_elgamal1985public} \cite{156_benaloh1994dense}. Somewhat homomorphic encryption schemes allow performing two types of operations limited number of times such as \cite{158_boneh2005evaluating} \cite{159_ishai2007evaluating}. Fully homomorphic encryption schemes allow  performing two type operations unlimited number of times such as \cite{160_van2010fully} \cite{161_brakerski2011fully} \cite{162_lopez2012fly}. The development of homomorphic encryption schemes is shown in Fig 7.
\begin{figure}[pos=htbp]
	\centering
	\includegraphics[scale=0.4]{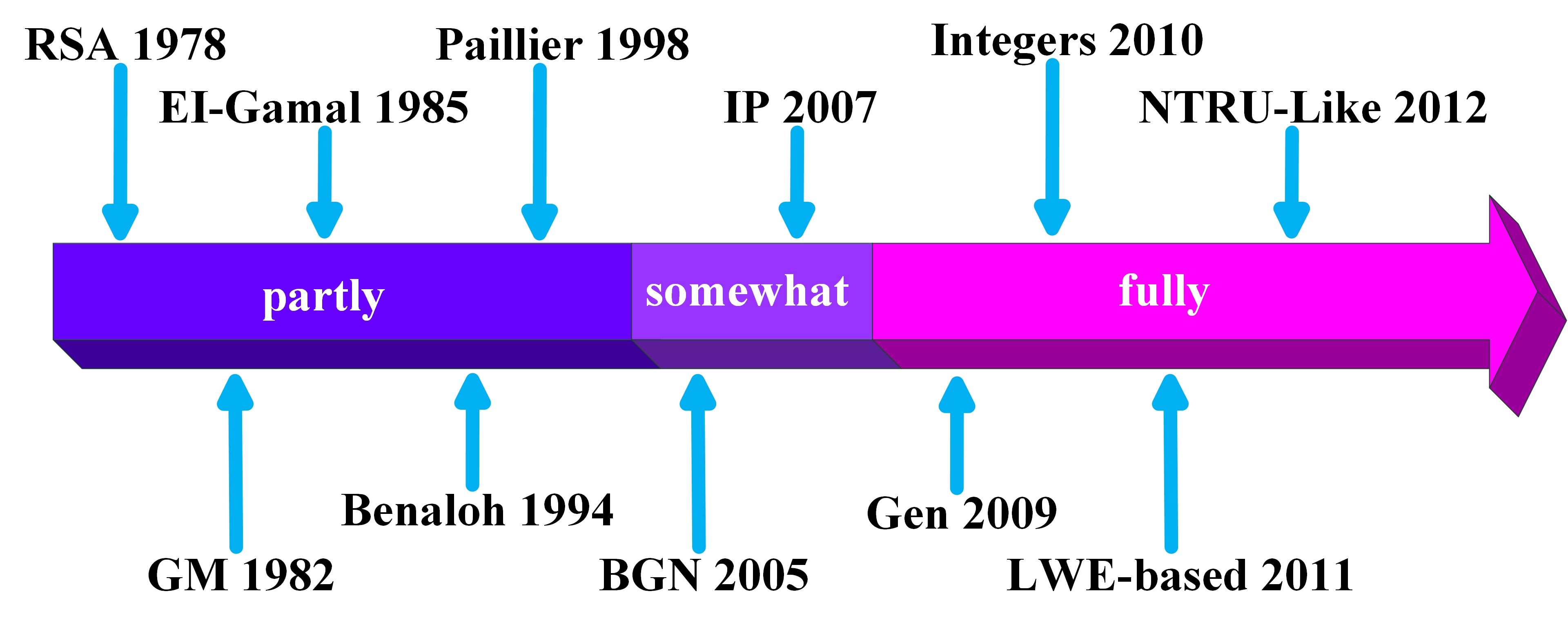}
	\subcaption*{Fig 7: Development of Homomorphic Encryption Schemes}
\end{figure}

\subsection{ Distributed Noise Generation}  

\par The distributed noise generation arises for two reasons. Firstly, the distributed noise generation can reduce the size of noise. In local differentially private mechanisms without homomorphic encryption schemes (secure multiple-party computation schemes), all participants add noise locally so that their privacy can be preserved, resulting in big noise. However, by homomorphic encryption schemes, each participant of these local differentially private mechanisms could locally add noise which is added up as one noise from appropriate noise distributions so that added noise of final results could be reduced.

\par Secondly, there is not a trusted aggregator in local differentially private mechanisms. Due to the absence of a trusted aggregator, no single participant should know overall added noise. Thus, noise added  to the final result should be generated in a distributed way.    

\subsubsection{ Distributed Laplace and Gaussian Noise}  

\par The distributed noise generation is that each participant generates partly noise such that aggregated noise obeys the Laplace distribution or the Gaussian distribution, which is enough to preserve differential privacy. This is possible due to infinite divisibility of the Laplace distribution and the Gaussian distribution \cite{163_kotz2012laplace}. In brief, noise from the Laplace distribution can be generated by multiple noise from the Gamma distribution, the Gaussian distribution or the Laplace distribution in a cooperatively way. And, noise from the Gaussian distribution can be generated by multiple noise from another Gaussian distributions in a cooperatively way.   

\par \textbf{Laplace noise generated by the Gamma distribution} Noise from the Laplace distribution can be generated by $2n$ noise from the Gamma distribution \cite{164_acs2011have}. Formally, 
\begin{eqnarray*}
	Lap(\mu,s) = \frac{\mu}{n} + \sum_{i=1}^{n}(G_i^1-G_i^2)
\end{eqnarray*} 
\par Where, $Lap(\mu,s)$ represents noise from the Laplace distribution with the mean of $\mu$ and the variance of $s$. $G_i^1$ and $G_i^2$ represent noise from the Gamma distribution with the density function as following
\begin{eqnarray*}
	pdf(G^1)=pdf(G^2)= pdf(x) = \frac{(1/2)^{1/n}}{\Gamma(1/n)} x^{1/n-1}e^{-x/s}
\end{eqnarray*} 

\par \textbf{Laplace noise generated by the Gaussian distribution} Noise from the Laplace distribution can be generated by $4$ noise from the Gaussian distribution \cite{165_rastogi2010differentially}. Namely, 
\begin{eqnarray*}
	Lap(0,s) &=& N_1(0,s/2)^2 + N_2(0,s/2)^2 \\ 
	& & - N_3(0,s/2)^2 - N_4(0,s/2)^2 
\end{eqnarray*} 
\par Here, $N_i(0,s/2)$ ($i=1,2,3,4$) represents noise from the Gaussian distribution with the mean of 0 and the variance of $s/2$.
\par In addition, noise from the Gaussian distribution can be generated by $n$ noise from another Gaussian distribution due to its infinite divisibility. Therefore,  
$N_i(0,s/2)$ for $i=1,2,3,4$ is equal to 
\begin{eqnarray*}
	N_i(0,s/2) = \sum_{j=1}^{n} N_j(0,s/2n) 
\end{eqnarray*}  
\par In a word, noise from the Laplace distribution can be generated by $4n$ noise from the Gaussian distribution. 

\par \textbf{Laplace noise generated by another Laplace distribution } Due to infinite divisibility of the Laplace distribution, noise from the Laplace distribution can be generated by noise from another Laplace distribution \cite{166_goryczka2013secure}. Formally,
\begin{eqnarray*}
	Lap(0,s) = \sqrt{B_{n-1}}\sum_{i=1}^{n}Lap_i(0,s)
\end{eqnarray*}  
\par $B_{n-1}$ represents noise from the Beta distribution with the probability density function  
\begin{eqnarray*}
	pdf(x) = (n-1)(1-x)^{n-2}
\end{eqnarray*}  
\par \textbf{Gaussian noise generated by other Gaussian distributions } Due to infinite divisibility of the Gaussian distribution, Gaussian noise could be simulated by $n$ Gaussian noise as following
\begin{eqnarray*}
	N(0,s) = \sum_{i=1}^{n} N_i(0,s/n) 
\end{eqnarray*} 

\par In a word, distributed Laplace noise could be used to construct distributed Laplace mechanisms, and distributed Gaussian noise could be used to construct distributed Gaussian mechanisms. 

\subsubsection{ Distributed Geometric Noise} 

\par Aforementioned distributed Laplace noise and distributed Gaussian noise are from continuous distributions. There are some application scenarios where outputs of differentially private mechanisms are expected to be integers not real numbers. For example, the issued query is "how many people have a certain property". For these application scenarios, Ghosh et al. propose the geometric mechanism \cite{173_ghosh2012universally} which perturbs outputs with noise from the geometric distribution.
\par \textbf{Geometric mechanism} A random mechanism is a $\epsilon$ geometric mechanism if for a query $q$ and a database $D$, the output of the mechanism is $q(D)+N$ where $N$ is an integer from a discrete geometric distribution with the following probability density function.
\begin{eqnarray*}
	pdf( N=x ) = \frac{1-\epsilon}{1+\epsilon}e^{|x|}
\end{eqnarray*}

\par If outputs of queries are limited to integer numbers in a bounded interval such as $[a, b]$, all outputs less than $a$ are remapped to $a$, and all outputs greater than $b$ are remapped to $b$. The approach is called truncated geometric mechanism which also satisfies requirements of $\epsilon$ geometric mechanism.   

\par \textbf{geometric noise generated by the P$\boldsymbol{\acute{o}}$lya distribution} Noise from the geometric distribution can be generated by $2n$ noise from the P$\boldsymbol{\acute{o}}$lya distribution \cite{45_goryczka2015comprehensive}. Formally, 
\begin{eqnarray*}
	N = \sum_{i=1}^n (X_i^1-X_i^2)
\end{eqnarray*} 
\par Where, $N$ represents noise from the geometric distribution. $X_i^1$ and $X_i^2$ are noise from the P$\acute{o}$lya distribution whit the probability density function as following
\begin{eqnarray*}
	pdf(X^1)=pdf(X^2)=pdf(x) =  \tbinom{1/n+x-1}{1/n-1} \epsilon^x(1-\epsilon)^{1/n}
\end{eqnarray*}   

\subsubsection{Distributed Noise Approximation} 

\par Various probability distributions are related to each other as shown in Fig 8 \cite{175} \cite{176}. Therefore, noise from the Gaussian distribution can be approximated by noise from the binomial distribution, and noise from the Laplace distribution(the exponential distribution) can be approximated by noise from the Poisson distribution.

\begin{figure*}[pos=htbp]
	\centering
	\includegraphics[scale=0.32]{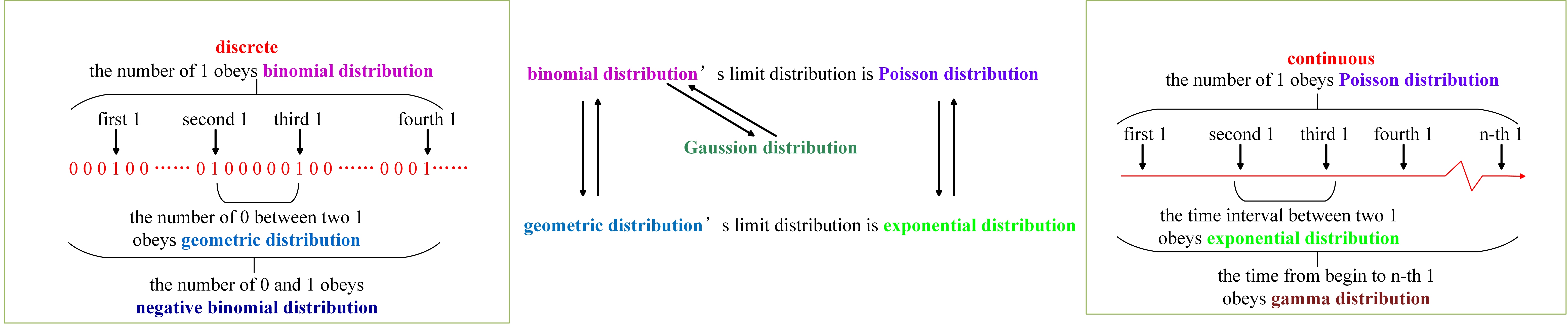}
	\subcaption*{Fig 8: Relationship among Various Probability Distributions}
\end{figure*}

\par Dwork et al. simulate noise from the exponential distribution by noise from the Poisson distribution, and simulate noise from the Gaussian distribution by noise from the binomial distribution \cite{178_dwork2006our}.

\par \textbf{Gaussian noise generated by the binomial distribution} First, simulate a public source of unbiased bits $c_1, c_2, \dots c_n$. In particular, each participant shares out two random bits by some secure multi-party computation method such as secrete sharing. These bits are represented by $b_1, b_2, \dots b_{2n}$. These bits can be transformed to unbiased bits $c_1, c_2, \dots c_n$ by deterministic extractor \cite{177_gabizon2006deterministic}. Second, each participant shares out a bit $a_i$, and $s_i$ is equal to $a_i\oplus c_i$. Third, $s_i$ is replaced by $2s-1$ such that 0 is remapped to -1 and 1 is still 1. At last, each participant sums these shares into a number to simulate noise from a binomial distribution. Simulated noise is used to approximate noise from the Gaussian distribution.  

\par \textbf{Laplace noise generated by the Poisson distribution} First, simulate a public source of unbiased bits as the first step in simulating Gaussian noise by binomial noise above. Second, noise from the Poisson distribution can be simulated based on a sequence of unbiased bits by circuits\cite{178_dwork2006our}. At last, noise from the Poisson distribution can be used to simulate noise from the Laplace distribution(exponential distribution).

\par Notably, these two methods need unbiased or biased bits which are generated by secure multi-party computation involving many participants. Therefore, these two methods have significant communication and computation overhead.

\subsubsection{Comparisons}  

\par In this subsection, the complexity of generating noise of every way is compared with each other, and the comparison is shown in Table 1. Notably, noise approximation methods introduced by Dwork et al. \cite{178_dwork2006our} are not included in Table 1. The reason is that these methods involve secure multi-party computation which dominants overhead not generating unbiased bits. The overhead of these two methods \cite{178_dwork2006our} is much higher than other methods.

\begin{table*}[pos=htbp]
	\footnotesize
	\centering
	\begin{tabular}{|c|c|c|c|c|}
		\hline 
		target noise&\tabincell{c}{tool \\ distribution}& \tabincell{c}{the number of \\ noise from tool \\ distribution per participant} & \tabincell{c}{the number of \\ unbiased bits per \\ tool distribution noise} & \tabincell{c}{the number of \\ unbiased bits in \\ total per participant} \\
		\hline
		\multirow{3}*{Laplace}&Gamma&2&at least 2 (2.54on average)& \tabincell{c}{at least 4 \\ (5.08on average )} \\
		\cline{2-5}
		&Gaussian&4& 1 on average& 4 on average \\
		\cline{2-5}
		&Laplace&1&1&1 \\	 
		\hline
		Gaussian&\tabincell{c}{other \\Gaussian}&1&1&1 \\
		\hline
		geometric&P$\acute{o}$lya&2 (1 Gamma and 1 Poisson)&\tabincell{c}{at least 2 for Gamma\\ (on average 2.54)  \\ and for Poisson \cite{179_devroye1986sample} }& at least 2 \\
		\hline
	\end{tabular}
	\subcaption*{Table 1: Complexity Comparison among Noise Generation Methods} 
\end{table*}

\par For Laplace noise generated by Gamma noise, simulating one noise from a Gamma distribution needs at least 2 noise from an uniform distribution \cite{167_ahrens1974computer}. If all values of $\alpha$ are equally probable for $Gamma(\alpha) = \int_{0}^{\infty}x^{\alpha-1}e^{-x} $, the average number of noise from an uniform distribution is equal to 2.54 \cite{45_goryczka2015comprehensive}.

\par For Laplace noise generated by Gaussian noise, simulating one noise from the Gaussian distribution needs one noise from an uniform distribution on average \cite{169_box1958note}. There are other ways which are more efficient on average but slow in the worst-case \cite{170_marsaglia1964convenient} \cite{171_marsaglia2000ziggurat}. 

\par For Laplace noise generated by other Laplace noise, simulating one noise from the Laplace distribution needs one noise from an uniform distribution. Specifically, the inverse cumulative distribution function is as following. Therefore, one noise from the Laplace distribution can be generated by one random number from the uniform distribution over the interval of $[0,1]$.     
\begin{eqnarray*}
	Lap(\mu,s) = \mu - sgn(x-0.5)*s*\ln(1-2|x-0.5|)
\end{eqnarray*}  
\par Here, $x$ is the random number from the uniform distribution over the interval of $[0,1]$. 

\par For Gaussian noise generated by other Gaussian noise, the complexity of generating one Gaussian noise is analyzed in generating Laplace noise by Gaussian noise above. 

\par For geometric noise generated by P$\acute{o}$lya noise, simulating one noise from the P$\acute{o}$lya distribution needs a number from the Gamma distribution and a number from the Poisson distribution \cite{174_johnson2005univariate}.

\subsection{Applications}  

\par The combination scheme between homomorphic encryption schemes and differentially private mechanisms is widely used in various data analysis tasks such as private record matching, private set operation, private machine learning, and so on.

\par Miran et al. combine differentially private mechanisms with homomorphic encryption schemes to achieve the best of both worlds \cite{46_kim2019secure}. They demonstrate their novel strategy by logistic regression over horizontally distributed datasets.  In particular, each participant locally computes gradient over his own datasets by the objective perturbation approach \cite{180_chaudhuri2011differentially}. Then, each participant chooses a random noise through distributed noise generation methods as mentioned before, which is added to local gradient such that aggregated noise preserves differential privacy. The perturbed local gradient is encrypted by homomorphic encryption schemes such that the aggregator can obtain encrypted global gradient by these encrypted local gradient. At last, a trusted third party decrypts that encrypted global gradient into plaintext and distributes plaintext to each participant for the next iteration.    

\par Bailey et al combine differentially private mechanisms with fully homomorphic encryption schemes to tackle the problem of private set intersection \cite{181_kacsmardifferentially}. In particular, they map elements to bins within a table by hashing-to-bins techniques \cite{182_groce2019cheaper} \cite{183_pinkas2015phasing} \cite{184_pinkas2018scalable}. Then, these plaintext is combined into a vector by a batching technique, and the vector is encrypted by fully homomorphic encryption scheme(BGV \cite{185_10.1145/2090236.2090262}). By homomorphic implementation of the randomized response algorithm, the server can compare his own elements with clients' elements in form of encrypted data in a way which is differential private. The batching process reduces overhead of homomorphic element comparisons through allowing processing of multiple plaintext in a batch. 

\par Acar et al. combine differentially private mechanisms with homomorphic encryption schemes to construct a distributed linear regression algorithm \cite{186_acar2017achieving}. The objective is to make sure that participants do not see each others’ data and cannot infer information about others from the final constructed statistical model. In particular, each participant computes his own statistic by functional mechanism \cite{187_zhang2012functional} and encrypts computed local statistics by homomorphic encryption schemes. These encrypted local statistics are sent to the next participant who add his own local statistic into encrypted data and send these encrypted data to the next participant. At last, a trusted participant decrypts encrypted data into plaintext and broadcasts the final result to all participants.

\par There are many other applications where differentially private mechanisms are combined with homomorphic encryption schemes to solve various problems. For example, Bao et al. combine differentially private mechanisms with homomorphic encryption schemes ( Boneh–Goh–Nissim scheme \cite{158_boneh2005evaluating}) to privately aggregate data generated by Smart Grid \cite{47_bao2015new}.

\subsection{Discussions}   

\par Fault tolerance or off-line of some participants needs to be considered carefully. One scheme is a fault tolerance scheme, which means that the scheme still can work even if some participants of the scheme are faulty.  In a distributed setting, there may be a very big number of participants. Thus, it is inevitable that some participants abort or go off-line. For example, communication channels are not available temporarily due to some accidents. However, some homomorphic encryption schemes have no ability to fault tolerance. Therefore, once one participant fails, the whole data aggregation protocol is not workable. 

\par The combination between differentially private mechanisms and homomorphic encryption schemes results in a more complex trade-off in the whole scheme. As we all know, there is a difficult trade-off between privacy and utility in differentially private mechanisms. As for homomorphic encryption schemes, the trade-off among communication overhead, length of ciphertext, support arithmetic operations, and so on is also difficult. When differentially private mechanisms are combined with homomorphic encryption schemes, there are more competitive goals to balance. Therefore, the combination makes the trade-off among various competitive goals extremely difficult.

\par Sensitivity of distributed differentially private mechanisms is hard to compute. In particular, sensitivity could be global sensitivity which is a property of queries. Therefore, it is relatively easy to compute by each participant locally although it wastes computation resources due to repeated calculation by each participant. However, global sensitivity always so big that utility of data is ruined totally if added noise is calibrated by global sensitivity. Other concepts of sensitivity such as record sensitivity \cite{150_huang2019efficient}, smooth sensitivity \cite{189_nissim2007smooth}, elastic sensitivity \cite{208_johnson2018towards}, and so on are dependent of data records although added noise is relatively smaller than global sensitivity if added noise is calibrated by these concepts of sensitivity. Dependence of data records means that communication overhead and computation overhead are significantly high. For example, smooth sensitivity is hard to be computed even if differentially private mechanisms are in a centralized setting. In the distributed setting, smooth sensitivity is even harder to be computed.   

\par Differentially private mechanisms also can be combined with other secure computation schemes such as secrete sharing \cite{190_bohler2020secure} \cite{192_eigner2015achieving} \cite{191_wu2016inherit}, garbled circuits \cite{193_yao1982protocols} \cite{194_yao1986generate}, trusted execution environment \cite{195_li2020efficient}. These different secure computation schemes have different properties in terms of communication overhead, the ability to fault tolerance, computation overhead, supported arithmetic operations, the complexity of schemes, and so on. Therefore, which secure computation scheme is appropriate depends on application scenarios.  



\section{Hardness Results Obtained by the One-way Function}   

\par The one-way function is the function that is easy to compute on every input, but hard to invert the image of any random input \cite{116_menezes1996handbook}. This description of the one-way function is two-fold. Firstly, the "easy" means that the one-way function can output results on every input in polynomial time. The "polynomial time" is another terminology in cryptography. In cryptography schemes, there is a secure parameter which is to quantify the secure strength of schemes, say $k$. The "polynomial time" means that running time is a polynomial formula of the secure parameter. For example, running time  $k^2+c$ and $k^{100}+k^{10}+ck+10$ are both polynomial running time where $c$ is a constant number. These polynomial time schemes are also called efficient schemes which refer to feasibility in terms of computing time. Secondly, the hardness from image to pre-image refers to that there is no polynomial time algorithm which can obtain pre-image for a given image.

\par The one-way function is usually based on some hard problems in the field of cryptography such as the discrete logarithm problem \cite{117_huang2019efficient}, big integer factorization problem, modular powers, and modular squaring. For example, the hardness of discrete logarithm problem is demonstrated in Fig 9. Specifically, the discrete logarithm problem is: for a prime number $n$ and a generator $g$ , given a number $a$, find a number $x$ such that $a = g^x\ mod\ n$. In this example shown in Fig 9, let $g=5,n=7,a=4$. The normal logarithm $\log_54$ can be computed by the binary search algorithm. In particular, select two numbers $x_1$ and $x_2$ such that $5^{x_1}<5^x<5^{x_2}$. Then let $x_3=(x_1+x_2)/2$. If $5^{x_3}<4$, then let $x_1=x_3$. Otherwise, let $x_2=x_3$. Compute $x_3=(x_1+x_2)/2$ again, and repeatedly do the same operations until $x_3$ is close to the value $\log_54$ enough(namely, $5^{x_3}$ is close to 4 enough). 

\begin{figure}[pos=htbp]
	\centering
	\includegraphics[scale=0.2]{./figure/discrete_logarithm_problem.jpg}
	\subcaption*{Fig 9: Comparison between Normal Logarithm and Discrete Logarithm}	
\end{figure}

\par However, for the discrete logarithm problem, the binary search algorithm does not work. In particular, for $a_1=4$, $x_1$ is 2($4=5^2\ mod\ 7$). However, for $a_2=1$, $x_2$ is 0($1=5^0\ mod\ 7$), and $a_3=2$, $x_3$ is 4($2=5^4\ mod\ 7$). As can be seen in Fig 9, although $a_2=1$ and $a_3=2$ are both less than $a_1=4$, $a_2=1$ lies on left of $a_1=4$, and $a_3=2$ lies on right of $a_1=4$. That is, the size relationship of $a$ fails to indicate the size relationship of $x$. Due to the failure of determining size relationship of $x$ by size relationship of $a$, the binary search algorithm cannot solve the discrete logarithm problem. Actually, there is no polynomial time algorithm which can solve the discrete logarithm problem right now if the prime number $n$ is sufficiently large.  

\par The discrete logarithm problem is appropriate to construct the one-way function \cite{147_shang2020full}. Firstly, for a given input $x$, $g^x\ mod\ n$ is easy to compute because there are only operations of calculating the exponential function and operations of calculating the mod function. Secondly, for a given $a$, it is hard to find $x$ such $a = g^x\ mod\ n$, which is interpreted above. That is, for discrete logarithm, it is hard to invert the image of any input.  

\par The one-way function is the security foundation of modern cryptography \cite{146_maurer2007information}. Specifically, cryptography schemes are constructed by a one-way function, which means that these schemes are based on difficult problems such as the discrete logarithm problem. The procedure of constructing cryptography schemes makes sure that the used difficult problem is solved if there is an attacker who can break one cryptography scheme. Because there is no efficient algorithm which can solve used difficult problem right now to the best knowledge, there is no attacker who can break cryptography schemes. The one-way function provides modern cryptography with a provable and firm mathematics foundation.

\subsection{Hardness Results} 

\par In differential privacy, an important research direction is to explore what is impossible to achieve in terms of utility \cite{42_vadhan2017complexity}. These theoretical upper boundaries and lower boundaries can avoid wasting engineers' time and energy when they try to design differentially private mechanisms for specialized applications. Until now, there have been some excellent papers which present hardness results for differential privacy based on the one-way function. These hardness results have a firmly mathematical foundation and clearly secure semantics due to the one-way function. 

\par Claims of hardness results in differential privacy usually are related to five factors including the size of data universal denoted by $|X|$, the size of query sets denoted by $|Q|$, the size of datasets denoted by $|D|$ which are sampled by a certain probability distribution over data universal, the boundary of error denoted by $e$, and running time of differentially private mechanisms denoted by $T$. Sometimes, the success probability of differentially private mechanisms denoted by $\beta$ is also mentioned. Claims are in the form of "For data universal with size $|X|$ and a query set with size $|Q|$, there is no polynomial time(or other claims of running time) differentially private mechanisms which can take a dataset with size $|D|$ as the input and output an accurate answers to every query in $|Q|$ up to the error of $e$ with high probability $\beta$".      

\par In this subsection, we focus on hardness results of differential privacy in three aspects including non-interactive query release, interactive query release, and synthetic data set generation. In the setting of non-interactive query release, differentially private mechanisms need to output answers to all possible queries in a query set with a given error boundary. While, in the setting of interactive query release, differentially private mechanisms just need to output answers to a certain number of queries in a query set not all queries in the query set although these queries to be answered might be chosen adaptively. The synthetic dataset is a specific dataset which is generated from a raw dataset such that the synthetic dataset is interchangeable with the raw dataset. In particular, the synthetic data can be used to accurately answer queries which can be accurately answered by the raw dataset.

\subsubsection{Non-interactive Query Release} 
\par Dwork et al. firstly build the connection between differentially private mechanisms and traitor-tracing schemes to demonstrate a hardness result of differential privacy in terms of non-interactive query release \cite{39_10.1145/1536414.1536467}. Traitor-tracing schemes are introduced by Benny et al. for protecting digital copyright \cite{118_chor1994tracing}. Roughly speaking, traitor-tracing schemes generate an encryption key as well as a tracing key for a sender and $n$ different decryption keys for $n$ different receivers. Traitor-tracing schemes make sure two guarantees. (1) Messages encrypted by the sender's encryption key can be decrypted by arbitrary receivers' decryption key. (2) If there is an efficient pirate decoder which is able to decrypt encrypted messages, traitor-tracing schemes can identify at least one receiver who contributes his decryption key to the pirate decoder no matter how receivers combine their keys to construct the pirate decoder.        

\par Traitor-tracing schemes consist of four algorithms including a setup algorithm, an encryption algorithm, a decryption algorithm, and a tracing algorithm \cite{148_zhandry2020new}. Specifically, the setup algorithm takes a security parameter $k$ as the input and outputs an encryption key for a sender, a tracing key for the tracing algorithm and different decryption keys for $n$ different receivers. The encryption algorithm takes messages and the encryption key as inputs and then outputs ciphertext. The decryption algorithm takes the ciphertext and a decryption key as inputs and then outputs the message which is encrypted by the sender. The tracing algorithm takes a decryption key(pirate decoder) as input and then outputs at least one receiver who contributes his decryption key to the pirate decoder.   

\par The amazing connection between differentially private mechanisms and traitor-tracing schemes is shown in Fig 10. In particular, there is a set of receivers who put their decryption keys into a dataset $D$ as data records. An sender encrypts possible messages into ciphertext which is used to generate queries. In Dwork's context, queries are statistical queries of form "what fraction of data records(decryption key) in the dataset satisfies the property specialized by a query (can be used to decrypt a given ciphertext)" \cite{119_10.1145/293347.293351}. Suppose there is a differential privacy mechanism(cracking algorithm) which takes the dataset $D$(some decryption keys) as the input and outputs a summary(pirate decoder) which can be used to answer queries(decrypt ciphertext) in the query set(set of all legal ciphertext). The differential privacy mechanism corresponds to a cracking algorithm of traitor-tracing schemes and the summary corresponds to a pirate decoder. The privacy attacker who tries to identify one record of the dataset corresponds to the tracing algorithm which tries to identify one decryption key of receivers. If the differential privacy mechanism keeps its privacy guarantee, that no algorithm can identify any one data record in the dataset by the summary, then the tracing algorithm cannot identify a receiver who contributes his decryption key to the pirate decoder. That is, secure traitor-tracing schemes do not exist. 

\begin{figure}[pos=htbp]
	\centering
	\includegraphics[scale=0.35]{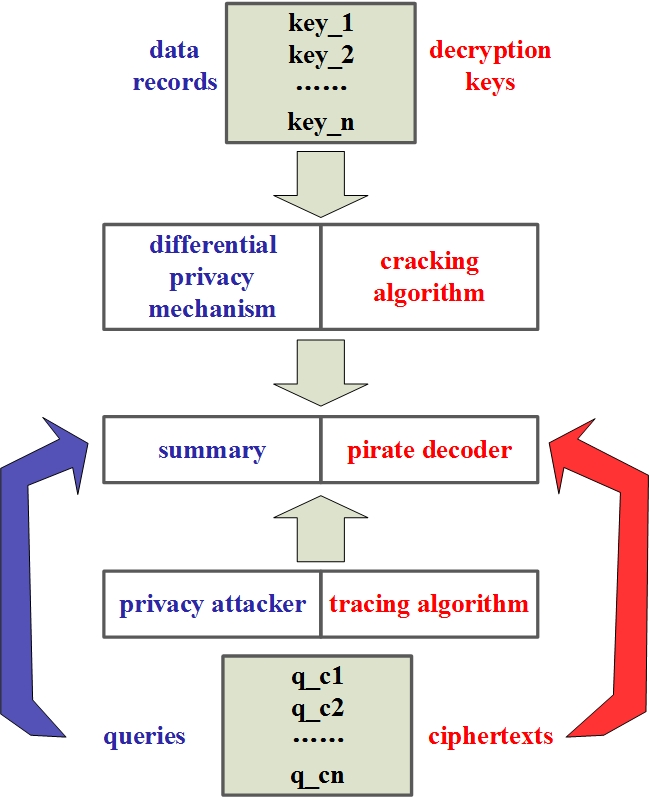}
	\subcaption*{Fig 10: Correspondence between Differential Privacy Mechanism and Traitor-tracing Scheme }
\end{figure}

\par Dwork et al. give a hardness result for differentially private mechanisms by the shown connection above in the context of non-interactive query release \cite{39_10.1145/1536414.1536467}. Specifically, for data universe $X = \{0,1\}^d(|X|=2^d)$ and a statistical query set $Q$ with a description length $ploy(n,d)(|Q|=2^{ploy(n,d)})$ where $n$ is the number of data records in the dataset $D$, there is no ($poly(n,d)$) polynomial time $(1,1/poly(n))$ differential privacy mechanism which can answer all queries in the query set $Q$ up to the error of $e=1/2-poly(n,d)$ with the probability $\beta=1-poly(n,d)$ if there exists a traitor-tracing scheme in which the length of a encryption key is $d$ and the length of ciphertext is $poly(n,d)$ where $n$ is the number of receivers. 

\par Lucas et al. give another hardness result for differentially private mechanisms based on the one-way function \cite{13_kowalczyk2018hardness}. For data universal $X$ with size $|X|=2^{2^{poly(\log\log k)}}$ and a statistical query set $|Q|$ with size $|Q|=2^k$, there is no polynomial time differential privacy mechanism which can take a dataset $D$ with size $n=poly(k)$ drawn from the data universal $X$ as the input and answer every query in the query set $Q$ up to an additive error of $\pm 1/3$ if the one-way function exists. Lucas et al. demonstrate that the existence of the one-way function implies the existence of traitor-tracing schemes by constructing a traitor-tracing scheme based on the one-way function. Therefore, the existence of the one-way function implies hardness results for differentially private mechanisms by the conclusion of Dwork \cite{39_10.1145/1536414.1536467}. In addition, compared with Dwork's traitor-tracing scheme, Lucas' traitor-tracing scheme supports a longer length for an encryption key so that data universal could be larger.

\subsubsection{Interactive Query Release}

\par Jonathan et al. give a hardness result for differentially private mechanisms in terms of interactive query release by stateful traitor-tracing schemes \cite{41_ullman2016answering}. Specifically, for data universe $X = \{0,1\}^d(|X|=2^d)$ and a statistical query set $Q$ with size $2^{n+o(1)}$ where $n$ is the number of data records in a dataset $D$, there is no ($poly(n,d)$) polynomial time $(1,1/poly(n))$ differential privacy mechanism  which takes the dataset $D$ as the input and returns an approximate answer to each query in the query set $Q$ up to the error of $\pm 0.49$ if the one-way function exists.

\par The connection between differentially private mechanisms and traitor-tracing schemes is shown by Dwork \cite{39_10.1145/1536414.1536467}. However, the traitor-tracing scheme used in Dwork's work is stateless traitor-tracing schemes. Jonathan et al. refine the connection between differentially private mechanisms and stateful traitor-tracing schemes \cite{121_kiayias2001crafty}, resulting in a hardness result of differentially private mechanisms in terms of interactive query release.    

\subsubsection{Synthetic Datasets}
\par Dwork et al. firstly build the connection between differentially private mechanisms and digital signature schemes to demonstrate hardness results of differential privacy in terms of synthetic dataset generation \cite{39_10.1145/1536414.1536467}. Digital signature schemes consist of three algorithms including a key generation algorithm, a signing algorithm, and a verification algorithm. Specifically, the key generation algorithm takes a security parameter as the input and outputs a secret key denoted by $sk$ and a public key denoted by $pk$. The signing algorithm takes a secret key together with a message $m$ as inputs and outputs a signature $\sigma=sign(sk,m)$. The verification algorithm takes a public key, a message $m$ together with a signature $\sigma$ as inputs and outputs an assertion that this signature is generated  by the message $m$ or not, namely $ver(sk,m,\sigma)= 1\ or\ 0$. If this signature is generated by the message, the verification algorithm outputs 1, otherwise 0. Roughly speaking, the digital signature schemes guarantee that there is no polynomial time cracking algorithm which can forge a signature $\sigma'$ for a given message $m'$ such that $ver(sk,m',\sigma')=1$.   

\par The connection between differentially private mechanisms and digital signature schemes is as follows. The signature algorithm randomly selects $n$ messages and generates their signatures. These pairs of a message and its signature are put into a dataset $D$. In Dwork's context, queries in query set $Q$ are statistical queries of form "what fraction of data records(pair of a message and its signature) in dataset can satisfy the property specialized by a query(satisfy verification algorithm)". Suppose there is a differential privacy mechanism which can take the dataset $D$ as the input and outputs a synthetic dataset $\hat{D}$ such that the synthetic dataset can be used to generate answers to queries in the query set. The synthetic dataset at least has one record $(\hat{m},\hat{\sigma})$. If queries in the query set obtain an accurate answer, the differential privacy mechanism successfully forges a pair of a message and its signature, which means that the differential privacy mechanism breaks the security of digital signature schemes. However, there is no polynomial time cracking algorithm for digital signature schemes at present. Due to the contradiction, supposing the existence of such differentially private mechanisms is wrong.

\par Dwork et al. give a hardness result for differentially private mechanisms in terms of generating synthetic dataset by the connection between differentially private mechanisms and digital signature schemes \cite{39_10.1145/1536414.1536467}. Specifically, for data universe $X = \{0,1\}^d(|X|=2^d)$ and a statistical query set $Q$ with a description length $ploy(n,d)(|Q|=2^{ploy(n,d)})$ where $n$ is the number of data records in the dataset $D$, there is no ($poly(n,d)$) polynomial time $(1,1/poly(n))$ differential privacy mechanism  which can take the dataset $D$ as the input and output a synthetic dataset $\hat{D}$ that can be used to answer all queries in the query set $Q$ up to the error of $e=1/3$ if there exists a secure digital signature scheme.   

\par Jonathan et al. give another hardness result for differentially private mechanisms in terms of generating synthetic dataset by probabilistically checkable proofs \cite{40_ullman2020pcps}. Specifically, for data universe $X = \{0,1\}^d(|X|=2^d)$ and a statistical query set $Q$ with a description length $ploy(n,d)(|Q|=2^{ploy(n,d)})$ where $n$ is the number of data records in a dataset $D$, there is no ($poly(n,d)$) polynomial time $(1,1/poly(n))$ differential privacy mechanism  which can take the dataset $D$ as the input and output a synthetic dataset $\hat{D}$ which can be used to answer all 2-way marginals queries in the query set $Q$ up to the error of $e=1/3$ if there exists the one-way function. Jonathan et al. show that statistical queries which are considered in Dwork's work \cite{39_10.1145/1536414.1536467} can be transformed into 2-way marginals queries by probabilistically checkable proofs \cite{120_bellare1993efficient}. In addition, the existence of the one-way function implies the existence of digital signature schemes. Therefore, Jonathan et al. reach their conclusion.

\begin{table*}[pos=htbp]
	\footnotesize
	\centering
	\begin{tabular}{|c|c|c|c|c|c|c|c|}
		\hline 
		reference & application&\tabincell{c}{size of \\ data \\ universal}& \tabincell{c}{size of \\ database } & \tabincell{c}{size of \\ query set} & \tabincell{c}{running \\ time}& \tabincell{c}{error \\ boundary}& \tabincell{c}{query \\ type} \\
		\hline
		Dwork \cite{39_10.1145/1536414.1536467} & \tabincell{c}{non-interactive\\ query release} & $2^d$ & $n$ &$2^{ploy(n,d)}$  & $poly(n,d)$ & \tabincell{c}{$1/2-$ \\ $poly(n,d)$} & \tabincell{c}{statistical \\ query}  \\
		\hline
		Lucas \cite{13_kowalczyk2018hardness} & \tabincell{c}{non-interactive\\ query release} & $2^{2^{poly(\log\log d)}}$ & $poly(d)$ & $2^d$  & $poly(d)$ & $\pm 1/3$ &\tabincell{c}{statistical \\ query} \\
		\hline
		Jonathan \cite{41_ullman2016answering} & \tabincell{c}{interactive\\ query release} & $2^d$ & $n$ &$2^{n+o(1)}$  & $poly(n,d)$  & $\pm 0.49$  & \tabincell{c}{statistical \\ query}  \\
		\hline
		Dwork \cite{39_10.1145/1536414.1536467} & \tabincell{c}{synthetic \\ data set} & $2^d$ & $n$ & $2^{ploy(n,d)}$ & $poly(n,d)$ &1/3 & \tabincell{c}{statistical \\ query}\\
		\hline
		Jonathan \cite{40_ullman2020pcps} & \tabincell{c}{synthetic \\ data set} & $2^d$ & $n$ & $2^{ploy(n,d)}$ & $poly(n,d)$ &1/3 & \tabincell{c}{2-way \\ marginals \\ queries}\\
		\hline
	\end{tabular}	
	\subcaption*{Table 2:  Comparison of Hardness Results} 
	\begin{tablenotes}
		\footnotesize
		\centering
		\item[] \textit{$d$ is the number of bits of a string which is used to present a data record. $n$ is the number of records in a database. $ploy()$ represents a polynomial with respect to its inputs. For example, $ploy(n,d)$ is a polynomial function with respect to $n$ and $d$.} 
		
	\end{tablenotes}
\end{table*}

\subsection{Discussions}

\par Two points of the  previous subsection are worth further explaining. Firstly, used queries of the previous subsection are somewhat complex and involve computing some cryptographic functionality. That is, these used queries are not nature queries in real applications. Therefore, these obtained hardness results still leave the possibility of designing efficient differentially private mechanisms for nature queries. Actually, there are papers discussing nature queries such as histograms(point functions)\cite{122_balcer2016efficient}, threshold functions \cite{123_dwork2010differential}, \cite{124_dwork2015pure}, \cite{125_beimel2013private} \cite{126_bun2015differentially}, conjunctions (marginals) \cite{127_thaler2012faster} \cite{128_chandrasekaran2014faster} \cite{129_dwork2015efficient}. 

\par Secondly, these obtained hardness results are under the constrain of polynomial time differentially private mechanisms. However, for a dataset $X=\{0,1\}^d$, if the $d$ is appropriately small, then exponential time differentially private mechanisms are feasible in real applications. Therefore, some researchers propose exponential time differentially private mechanisms for the relatively small dataset for a better accuracy. For example, Barak et al. give an exponential time differentially private mechanism for generating synthetic datasets \cite{130_barak2007privacy}. Blum et al. give another exponential time differential private mechanism( with subsequent developments in \cite{39_10.1145/1536414.1536467} \cite{133_dwork2010boosting} \cite{138_gupta2012iterative} \cite{135_hardt2012simple}) for answering more than $n^2$ queries \cite{136_10.1145/2450142.2450148}.   

\par The key to obtain hardness results of differential privacy through cryptography is to find the correspondence between differentially private mechanisms' components and cryptography schemes' components like \cite{39_10.1145/1536414.1536467}. In the correspondence relationship, the differentially private mechanisms play the role of a cracking algorithm in cryptography such that the existence of differentially private mechanisms implies the existence of a cracking algorithm of cryptography, resulting in a contradiction. Due to the contradiction, the conclusion of no such differentially private mechanisms can be achieved.    

\par Except for the one-way function, there are other approaches which are possibly used to obtain hardness results for differentially private mechanisms. The first approach is probably approximately correct(PAC) learning. A series of works have shown that an efficient PAC learning algorithm can be used to obtain an efficient differentially private mechanism \cite{136_10.1145/2450142.2450148} \cite{137_gupta2013privately} \cite{138_gupta2012iterative} \cite{139_hardt2012private}. Therefore, if there is a hardness result for PAC learning, intuitively, there might be a corresponding hardness result for differentially private mechanisms. The second approach is information theory. Notably, a privacy attacker in information theory generally has infinite computing power, which is different from cryptography where a privacy attacker only has polynomial computing power. There are some excellent papers which give hardness results for differential privacy based on information theory such as \cite{140_nikolov2013geometry} \cite{141_muthukrishnan2012optimal} \cite{142_kasiviswanathan2010price} \cite{143_matouvsek2020factorization} \cite{144_beimel2010bounds} \cite{145_kairouz2015composition}. These hardness results obtained by information theory for differential privacy are discussed more deeply in Vadhan's tutorial \cite{42_vadhan2017complexity}.

\section{Future Direction} 
\par Although differential privacy has developed for many years, it is still a thriving research topic. Due to wide and successful applications in various fields, there arises a lot of new problems to be solved for differential privacy. 

\subsection{Distributed Sensitivity Calculation}
\par Sensitivity is a necessary parameter for each participant in local differential privacy models. Specifically, in local differential privacy models, each participant perturbs his data locally such that privacy can be preserved even if an aggregator might attempt to infer participants' data, leading to large noise in the final aggregation result. To reduce the size of noise, secure multi-party computation methods are introduced to local differential privacy models so that each participant can generate partly noise whose summary can preserve privacy of the final result. Therefore, each participant needs to know sensitivity of queries because sensitivity is a key parameter of local noise distributions. For example, to generate noise from the Laplace distribution in a distributed way by the Gamma distribution, each participant generates partly noise from the Gamma distribution with density function as following
\begin{eqnarray*}
	pdf(x) = \frac{(1/2)^{1/n}}{\Gamma(1/n)} x^{1/n-1}e^{-x/s}
\end{eqnarray*}    
\par Here, the parameter $s=2(\Delta D/\epsilon)^2$ is the variance of the Laplace distribution. So, all participants need to know sensitivity $\Delta D$.  

\par How to calculate sensitivity in a distributed way needs to further explore. The first concept of sensitivity is global sensitivity which is a property of queries. Therefore, global sensitivity can be calculated by each participant locally according to issued queries. However, if noise is calibrated by global sensitivity, the added noise is so big that utility of data is totally ruined in many real-word applications. To that end, many concepts of database-based sensitivity such as restricted sensitivity \cite{225_blocki2013differentially} are proposed. Due to the absence of a trusted aggregator, database-based sensitivity needs to be calculated in a collaborative way in local differential privacy model. Unfortunately, the process of calculating database-based sensitivity involves data records which differentially private mechanisms are intended to protect. It is possible to leak privacy information of data in the process of calculating database-based sensitivity. How to privately calculate database-based sensitivity needs to further explore. In addition, some concepts of database-based sensitivity are hard to compute even if this calculation is in the centralized differential privacy model such as smooth sensitivity \cite{189_nissim2007smooth}. In a distributed setting, this calculation is harder. So, in a word, how to privately and efficiently calculate database-based sensitivity in a distributed setting deserves further study.    

\subsection{New Cryptography Primitive}

\par The privacy guarantee of cryptography schemes could be relaxed when they are combined with differentially private mechanisms for privacy preservation. The reason is two-fold. Firstly, although cryptography schemes can guarantee that nothing can be learned except finial results, the differentially private mechanisms allow leaking a few information which is strictly controlled by the privacy parameter $\epsilon$. Therefore, when these two kinds of mechanisms are combined with each other, a few information is legally leaked even if cryptography schemes leak nothing. So, the privacy guarantee of cryptography schemes could be relaxed to a new guarantee. For example, cryptography schemes could be allowed to leak a few information which is controlled by the privacy parameter.

\par Secondly, relaxing privacy guarantees of cryptography schemes can reduce computing overhead of combination schemes between differentially private mechanisms and cryptography schemes. In particular, the cryptography schemes are based on some cryptographic primitive operations whose computing overhead is significant such as the operation of encryption. Prohibitively high computing overhead stops applications of combined schemes. Fortunately, some papers point out that computing overhead can be dramatically reduced if the privacy guarantee of cryptography schemes can be relaxed. For example, Mazloom et al. come up with an idea: `` instead of claiming that the servers(attackers) learn nothing about the input values, they claim that what servers(attackers) do learn from the computation process preserves the differential privacy of the input." \cite{196_mazloom2018secure}. Wagh et al. introduce differential privacy into oblivious RAM \cite{197_wagh2018differentially}. By theoretical analyses, simulations, and real-world implementation, they demonstrate that differential privacy enhances the performance of oblivious RAM mechanisms while providing rigorous privacy guarantees.  

\par In brief, the privacy guarantee of cryptography schemes is not necessary to be that nothing can be learned, and relaxing privacy guarantee can reduce computing overhead of combination schemes between differentially private mechanisms and cryptography schemes. So, in order to pave the way for wider applications of differentially private mechanisms, it is worth exploring new cryptography primitives which allow leaking limited information while preserving differential privacy for inputs values.     

\subsection{ Improving Utility } 

\par Improving utility of differentially private mechanisms promotes the development of differential privacy. For example, to improve data utility, various concepts of sensitivity are proposed. These concepts make it possible to add database-based noise so that added noise could be reduced. Another example is the introduction of secure multi-party computation. Specifically, each participant of local differentially private mechanisms perturbs his data locally, resulting in big noise in the final result. The introduction of secure multi-party computation enables each participant of local differential privacy model to add smaller noise while preserving privacy of the final result.  

\par At present, many technologies including cryptography schemes can improve utility of differential privacy. As mentioned before, shuffling by anonymous communication can improve data utility of differential privacy \cite{37_cheu2019distributed} \cite{228_balle2019privacy}. There are other technologies which can be used to improve utility of differentially private mechanisms including sub-sampling \cite{229_balle2020privacy}, iteration \cite{230_feldman2018privacy} \cite{231_asoodeh2020privacy}, and so on. 

\par The differentially private mechanisms are more frequently used in more complex data analysis tasks such as machining learning. Compared with traditional data analysis tasks including queries release and statistical estimation, these more complex data analysis tasks require for higher utility. In addition, sophisticated data analysis technologies bring huge profits to commercial companies. Therefore, these commercial companies take advantage of their own large datasets to support complex data analysis tasks for improving their services. However, these used datasets have a lot of privacy information of individuals. It is great significant for preserving personal privacy to improve utility of differentially private mechanisms so that these mechanisms can work in complex data analysis tasks.   

\section{Conclusion}
\par Due to the wide and successful applications of data analysis technologies, many institutions have accumulated rich data to improve their services. However, personal data are collected, stored, and shared among various institutions for research goals or commercial goals, leading to increasing concern about personal privacy. Especially, there are many data leakage events. For example, all 6.5 million Israeli voters are exposed \cite{233_victor2020personal}.  

\par Differential privacy is a promising tool to handle the problem of privacy preservation so it has drawn significant attention. In this survey, we elaborate that how to improving utility of differential privacy through cryptography and what is impossible to achieve for utility of differential privacy from the lens of cryptography. Specifically, we analyze the utility amplification effect of anonymous communication. Then, we investigate the combination between differentially private mechanisms and homomorphic encryption schemes for purpose of improving utility. Next, we present hardness results about what is impossible for differentially private mechanisms' utility. These hardness results are obtained through the one-way function.   

\par Differential privacy benefited from cryptography in past and will benefit from new developments of cryptography in the future. We are motivated to make the survey for summarizing the state-of-the-art and promoting further developments of differential privacy.

\bibliographystyle{cas-model2-names}
\bibliography{acmart}

\begin{thebibliography}{146}
\expandafter\ifx\csname natexlab\endcsname\relax\def\natexlab#1{#1}\fi
\providecommand{\url}[1]{\texttt{#1}}
\providecommand{\href}[2]{#2}
\providecommand{\path}[1]{#1}
\providecommand{\DOIprefix}{doi:}
\providecommand{\ArXivprefix}{arXiv:}
\providecommand{\URLprefix}{URL: }
\providecommand{\Pubmedprefix}{pmid:}
\providecommand{\doi}[1]{\href{http://dx.doi.org/#1}{\path{#1}}}
\providecommand{\Pubmed}[1]{\href{pmid:#1}{\path{#1}}}
\providecommand{\bibinfo}[2]{#2}
\ifx\xfnm\relax \def\xfnm[#1]{\unskip,\space#1}\fi
\bibitem[{Abadi et~al.(2016)Abadi, Chu, Goodfellow, McMahan, Mironov, Talwar
  and Zhang}]{236_abadi2016deep}
\bibinfo{author}{Abadi, M.}, \bibinfo{author}{Chu, A.},
  \bibinfo{author}{Goodfellow, I.}, \bibinfo{author}{McMahan, H.B.},
  \bibinfo{author}{Mironov, I.}, \bibinfo{author}{Talwar, K.},
  \bibinfo{author}{Zhang, L.}, \bibinfo{year}{2016}.
\newblock \bibinfo{title}{Deep learning with differential privacy}, in:
  \bibinfo{booktitle}{Proceedings of the 2016 ACM SIGSAC Conference on Computer
  and Communications Security}, pp. \bibinfo{pages}{308--318}.
\bibitem[{Acar et~al.(2018)Acar, Aksu, Uluagac and Conti}]{43_acar2018survey}
\bibinfo{author}{Acar, A.}, \bibinfo{author}{Aksu, H.},
  \bibinfo{author}{Uluagac, A.S.}, \bibinfo{author}{Conti, M.},
  \bibinfo{year}{2018}.
\newblock \bibinfo{title}{A survey on homomorphic encryption schemes: Theory
  and implementation}.
\newblock \bibinfo{journal}{ACM Computing Surveys (CSUR)} \bibinfo{volume}{51},
  \bibinfo{pages}{1--35}.
\bibitem[{Acar et~al.(2017)Acar, Celik, Aksu, Uluagac and
  McDaniel}]{186_acar2017achieving}
\bibinfo{author}{Acar, A.}, \bibinfo{author}{Celik, Z.B.},
  \bibinfo{author}{Aksu, H.}, \bibinfo{author}{Uluagac, A.S.},
  \bibinfo{author}{McDaniel, P.}, \bibinfo{year}{2017}.
\newblock \bibinfo{title}{Achieving secure and differentially private
  computations in multiparty settings}, in: \bibinfo{booktitle}{2017 IEEE
  Symposium on Privacy-Aware Computing (PAC)}, \bibinfo{organization}{IEEE}.
  pp. \bibinfo{pages}{49--59}.
\bibitem[{{\'A}cs and Castelluccia(2011)}]{164_acs2011have}
\bibinfo{author}{{\'A}cs, G.}, \bibinfo{author}{Castelluccia, C.},
  \bibinfo{year}{2011}.
\newblock \bibinfo{title}{I have a dream!(differentially private smart
  metering)}, in: \bibinfo{booktitle}{International Workshop on Information
  Hiding}, \bibinfo{organization}{Springer}. pp. \bibinfo{pages}{118--132}.
\bibitem[{Ahrens and Dieter(1974)}]{167_ahrens1974computer}
\bibinfo{author}{Ahrens, J.H.}, \bibinfo{author}{Dieter, U.},
  \bibinfo{year}{1974}.
\newblock \bibinfo{title}{Computer methods for sampling from gamma, beta,
  poisson and bionomial distributions}.
\newblock \bibinfo{journal}{Computing} \bibinfo{volume}{12},
  \bibinfo{pages}{223--246}.
\bibitem[{Alvim et~al.(2018)Alvim, Chatzikokolakis, Palamidessi and
  Pazii}]{104_alvim2018local}
\bibinfo{author}{Alvim, M.}, \bibinfo{author}{Chatzikokolakis, K.},
  \bibinfo{author}{Palamidessi, C.}, \bibinfo{author}{Pazii, A.},
  \bibinfo{year}{2018}.
\newblock \bibinfo{title}{Local differential privacy on metric spaces:
  optimizing the trade-off with utility}, in: \bibinfo{booktitle}{2018 IEEE
  31st Computer Security Foundations Symposium (CSF)},
  \bibinfo{organization}{IEEE}. pp. \bibinfo{pages}{262--267}.
\bibitem[{anonymous(2020)}]{175}
\bibinfo{author}{anonymous}, \bibinfo{year}{2020}.
\newblock \bibinfo{title}{Univariate distribution relationships}.
\newblock \bibinfo{howpublished}{[EB/OL]}.
\newblock
  \bibinfo{note}{\url{http://www.math.wm.edu/~leemis/chart/UDR/UDR.html}}.
\bibitem[{Asoodeh et~al.(2020)Asoodeh, Diaz and
  Calmon}]{231_asoodeh2020privacy}
\bibinfo{author}{Asoodeh, S.}, \bibinfo{author}{Diaz, M.},
  \bibinfo{author}{Calmon, F.P.}, \bibinfo{year}{2020}.
\newblock \bibinfo{title}{Privacy amplification of iterative algorithms via
  contraction coefficients}.
\newblock \bibinfo{journal}{arXiv preprint arXiv:2001.06546} .
\bibitem[{Attasena et~al.(2017)Attasena, Darmont and
  Harbi}]{149_attasena2017secret}
\bibinfo{author}{Attasena, V.}, \bibinfo{author}{Darmont, J.},
  \bibinfo{author}{Harbi, N.}, \bibinfo{year}{2017}.
\newblock \bibinfo{title}{Secret sharing for cloud data security: a survey}.
\newblock \bibinfo{journal}{The VLDB Journal} \bibinfo{volume}{26},
  \bibinfo{pages}{657--681}.
\bibitem[{Balcer and Vadhan(2016)}]{122_balcer2016efficient}
\bibinfo{author}{Balcer, V.}, \bibinfo{author}{Vadhan, S.},
  \bibinfo{year}{2016}.
\newblock \bibinfo{title}{Efficient algorithms for differentially private
  histograms with worst-case accuracy over large domains}.
\newblock \bibinfo{journal}{Manuscript, July} .
\bibitem[{Balle et~al.(2018)Balle, Barthe and Gaboardi}]{109_balle2018privacy}
\bibinfo{author}{Balle, B.}, \bibinfo{author}{Barthe, G.},
  \bibinfo{author}{Gaboardi, M.}, \bibinfo{year}{2018}.
\newblock \bibinfo{title}{Privacy amplification by subsampling: Tight analyses
  via couplings and divergences}, in: \bibinfo{booktitle}{Advances in Neural
  Information Processing Systems}, pp. \bibinfo{pages}{6277--6287}.
\bibitem[{Balle et~al.(2020)Balle, Barthe and Gaboardi}]{229_balle2020privacy}
\bibinfo{author}{Balle, B.}, \bibinfo{author}{Barthe, G.},
  \bibinfo{author}{Gaboardi, M.}, \bibinfo{year}{2020}.
\newblock \bibinfo{title}{Privacy profiles and amplification by subsampling}.
\newblock \bibinfo{journal}{Journal of Privacy and Confidentiality}
  \bibinfo{volume}{10}.
\bibitem[{Balle et~al.(2019)Balle, Barthe, Gaboardi and
  Geumlek}]{228_balle2019privacy}
\bibinfo{author}{Balle, B.}, \bibinfo{author}{Barthe, G.},
  \bibinfo{author}{Gaboardi, M.}, \bibinfo{author}{Geumlek, J.},
  \bibinfo{year}{2019}.
\newblock \bibinfo{title}{Privacy amplification by mixing and diffusion
  mechanisms}, in: \bibinfo{booktitle}{Advances in Neural Information
  Processing Systems}, pp. \bibinfo{pages}{13298--13308}.
\bibitem[{Bao and Lu(2015)}]{47_bao2015new}
\bibinfo{author}{Bao, H.}, \bibinfo{author}{Lu, R.}, \bibinfo{year}{2015}.
\newblock \bibinfo{title}{A new differentially private data aggregation with
  fault tolerance for smart grid communications}.
\newblock \bibinfo{journal}{IEEE Internet of Things Journal}
  \bibinfo{volume}{2}, \bibinfo{pages}{248--258}.
\bibitem[{Barak et~al.(2007)Barak, Chaudhuri, Dwork, Kale, McSherry and
  Talwar}]{130_barak2007privacy}
\bibinfo{author}{Barak, B.}, \bibinfo{author}{Chaudhuri, K.},
  \bibinfo{author}{Dwork, C.}, \bibinfo{author}{Kale, S.},
  \bibinfo{author}{McSherry, F.}, \bibinfo{author}{Talwar, K.},
  \bibinfo{year}{2007}.
\newblock \bibinfo{title}{Privacy, accuracy, and consistency too: a holistic
  solution to contingency table release}, in: \bibinfo{booktitle}{Proceedings
  of the twenty-sixth ACM SIGMOD-SIGACT-SIGART symposium on Principles of
  database systems}, pp. \bibinfo{pages}{273--282}.
\bibitem[{Beimel et~al.(2010)Beimel, Kasiviswanathan and
  Nissim}]{144_beimel2010bounds}
\bibinfo{author}{Beimel, A.}, \bibinfo{author}{Kasiviswanathan, S.P.},
  \bibinfo{author}{Nissim, K.}, \bibinfo{year}{2010}.
\newblock \bibinfo{title}{Bounds on the sample complexity for private learning
  and private data release}, in: \bibinfo{booktitle}{Theory of Cryptography
  Conference}, \bibinfo{organization}{Springer}. pp. \bibinfo{pages}{437--454}.
\bibitem[{Beimel et~al.(2013)Beimel, Nissim and
  Stemmer}]{125_beimel2013private}
\bibinfo{author}{Beimel, A.}, \bibinfo{author}{Nissim, K.},
  \bibinfo{author}{Stemmer, U.}, \bibinfo{year}{2013}.
\newblock \bibinfo{title}{Private learning and sanitization: Pure vs.
  approximate differential privacy}, in: \bibinfo{booktitle}{Approximation,
  Randomization, and Combinatorial Optimization. Algorithms and Techniques}.
  \bibinfo{publisher}{Springer}, pp. \bibinfo{pages}{363--378}.
\bibitem[{Bellare et~al.(1993)Bellare, Goldwasser, Lund and
  Russell}]{120_bellare1993efficient}
\bibinfo{author}{Bellare, M.}, \bibinfo{author}{Goldwasser, S.},
  \bibinfo{author}{Lund, C.}, \bibinfo{author}{Russell, A.},
  \bibinfo{year}{1993}.
\newblock \bibinfo{title}{Efficient probabilistically checkable proofs and
  applications to approximations}, in: \bibinfo{booktitle}{Proceedings of the
  twenty-fifth annual ACM symposium on Theory of computing}, pp.
  \bibinfo{pages}{294--304}.
\bibitem[{Benaloh(1994)}]{156_benaloh1994dense}
\bibinfo{author}{Benaloh, J.}, \bibinfo{year}{1994}.
\newblock \bibinfo{title}{Dense probabilistic encryption}, in:
  \bibinfo{booktitle}{Proceedings of the workshop on selected areas of
  cryptography}, pp. \bibinfo{pages}{120--128}.
\bibitem[{Bittau et~al.(2017)Bittau, Erlingsson, Maniatis, Mironov,
  Raghunathan, Lie, Rudominer, Kode, Tinnes and Seefeld}]{36_bittau2017prochlo}
\bibinfo{author}{Bittau, A.}, \bibinfo{author}{Erlingsson, {\'U}.},
  \bibinfo{author}{Maniatis, P.}, \bibinfo{author}{Mironov, I.},
  \bibinfo{author}{Raghunathan, A.}, \bibinfo{author}{Lie, D.},
  \bibinfo{author}{Rudominer, M.}, \bibinfo{author}{Kode, U.},
  \bibinfo{author}{Tinnes, J.}, \bibinfo{author}{Seefeld, B.},
  \bibinfo{year}{2017}.
\newblock \bibinfo{title}{Prochlo: Strong privacy for analytics in the crowd},
  in: \bibinfo{booktitle}{Proceedings of the 26th Symposium on Operating
  Systems Principles}, pp. \bibinfo{pages}{441--459}.
\bibitem[{Blocki et~al.(2013)Blocki, Blum, Datta and
  Sheffet}]{225_blocki2013differentially}
\bibinfo{author}{Blocki, J.}, \bibinfo{author}{Blum, A.},
  \bibinfo{author}{Datta, A.}, \bibinfo{author}{Sheffet, O.},
  \bibinfo{year}{2013}.
\newblock \bibinfo{title}{Differentially private data analysis of social
  networks via restricted sensitivity}, in: \bibinfo{booktitle}{Proceedings of
  the 4th conference on Innovations in Theoretical Computer Science}, pp.
  \bibinfo{pages}{87--96}.
\bibitem[{Blum et~al.(2005)Blum, Dwork, McSherry and
  Nissim}]{212_blum2005practical}
\bibinfo{author}{Blum, A.}, \bibinfo{author}{Dwork, C.},
  \bibinfo{author}{McSherry, F.}, \bibinfo{author}{Nissim, K.},
  \bibinfo{year}{2005}.
\newblock \bibinfo{title}{Practical privacy: the sulq framework}, in:
  \bibinfo{booktitle}{Proceedings of the twenty-fourth ACM SIGMOD-SIGACT-SIGART
  symposium on Principles of database systems}, pp. \bibinfo{pages}{128--138}.
\bibitem[{Blum et~al.(2013)Blum, Ligett and Roth}]{136_10.1145/2450142.2450148}
\bibinfo{author}{Blum, A.}, \bibinfo{author}{Ligett, K.},
  \bibinfo{author}{Roth, A.}, \bibinfo{year}{2013}.
\newblock \bibinfo{title}{A learning theory approach to noninteractive database
  privacy}.
\newblock \bibinfo{journal}{J. ACM} \bibinfo{volume}{60}.
\bibitem[{B{\"o}hler and Kerschbaum(2020)}]{190_bohler2020secure}
\bibinfo{author}{B{\"o}hler, J.}, \bibinfo{author}{Kerschbaum, F.},
  \bibinfo{year}{2020}.
\newblock \bibinfo{title}{Secure multi-party computation of differentially
  private median}, in: \bibinfo{booktitle}{29th $\{$USENIX$\}$ Security
  Symposium ($\{$USENIX$\}$ Security 20)}, pp. \bibinfo{pages}{2147--2164}.
\bibitem[{Boneh et~al.(2005)Boneh, Goh and Nissim}]{158_boneh2005evaluating}
\bibinfo{author}{Boneh, D.}, \bibinfo{author}{Goh, E.J.},
  \bibinfo{author}{Nissim, K.}, \bibinfo{year}{2005}.
\newblock \bibinfo{title}{Evaluating 2-dnf formulas on ciphertexts}, in:
  \bibinfo{booktitle}{Theory of cryptography conference},
  \bibinfo{organization}{Springer}. pp. \bibinfo{pages}{325--341}.
\bibitem[{Box(1958)}]{169_box1958note}
\bibinfo{author}{Box, G.E.}, \bibinfo{year}{1958}.
\newblock \bibinfo{title}{A note on the generation of random normal deviates}.
\newblock \bibinfo{journal}{Ann. Math. Stat.} \bibinfo{volume}{29},
  \bibinfo{pages}{610--611}.
\bibitem[{Brakerski et~al.(2012)Brakerski, Gentry and
  Vaikuntanathan}]{185_10.1145/2090236.2090262}
\bibinfo{author}{Brakerski, Z.}, \bibinfo{author}{Gentry, C.},
  \bibinfo{author}{Vaikuntanathan, V.}, \bibinfo{year}{2012}.
\newblock \bibinfo{title}{(leveled) fully homomorphic encryption without
  bootstrapping}, \bibinfo{publisher}{Association for Computing Machinery},
  \bibinfo{address}{New York, NY, USA}.
\bibitem[{Brakerski and Vaikuntanathan(2011)}]{161_brakerski2011fully}
\bibinfo{author}{Brakerski, Z.}, \bibinfo{author}{Vaikuntanathan, V.},
  \bibinfo{year}{2011}.
\newblock \bibinfo{title}{Fully homomorphic encryption from ring-lwe and
  security for key dependent messages}, in: \bibinfo{booktitle}{Annual
  cryptology conference}, \bibinfo{organization}{Springer}. pp.
  \bibinfo{pages}{505--524}.
\bibitem[{Bun et~al.(2015)Bun, Nissim, Stemmer and
  Vadhan}]{126_bun2015differentially}
\bibinfo{author}{Bun, M.}, \bibinfo{author}{Nissim, K.},
  \bibinfo{author}{Stemmer, U.}, \bibinfo{author}{Vadhan, S.},
  \bibinfo{year}{2015}.
\newblock \bibinfo{title}{Differentially private release and learning of
  threshold functions}, in: \bibinfo{booktitle}{2015 IEEE 56th Annual Symposium
  on Foundations of Computer Science}, \bibinfo{organization}{IEEE}. pp.
  \bibinfo{pages}{634--649}.
\bibitem[{Cao et~al.(2018)Cao, Yoshikawa, Xiao and
  Xiong}]{206_cao2018quantifying}
\bibinfo{author}{Cao, Y.}, \bibinfo{author}{Yoshikawa, M.},
  \bibinfo{author}{Xiao, Y.}, \bibinfo{author}{Xiong, L.},
  \bibinfo{year}{2018}.
\newblock \bibinfo{title}{Quantifying differential privacy in continuous data
  release under temporal correlations}.
\newblock \bibinfo{journal}{IEEE transactions on knowledge and data
  engineering} \bibinfo{volume}{31}, \bibinfo{pages}{1281--1295}.
\bibitem[{Chandrasekaran et~al.(2014)Chandrasekaran, Thaler, Ullman and
  Wan}]{128_chandrasekaran2014faster}
\bibinfo{author}{Chandrasekaran, K.}, \bibinfo{author}{Thaler, J.},
  \bibinfo{author}{Ullman, J.}, \bibinfo{author}{Wan, A.},
  \bibinfo{year}{2014}.
\newblock \bibinfo{title}{Faster private release of marginals on small
  databases}, in: \bibinfo{booktitle}{Proceedings of the 5th conference on
  Innovations in theoretical computer science}, pp. \bibinfo{pages}{387--402}.
\bibitem[{Chaudhuri et~al.(2011)Chaudhuri, Monteleoni and
  Sarwate}]{180_chaudhuri2011differentially}
\bibinfo{author}{Chaudhuri, K.}, \bibinfo{author}{Monteleoni, C.},
  \bibinfo{author}{Sarwate, A.D.}, \bibinfo{year}{2011}.
\newblock \bibinfo{title}{Differentially private empirical risk minimization.}
\newblock \bibinfo{journal}{Journal of Machine Learning Research}
  \bibinfo{volume}{12}.
\bibitem[{Chaum(1981)}]{100_chaum1981untraceable}
\bibinfo{author}{Chaum, D.L.}, \bibinfo{year}{1981}.
\newblock \bibinfo{title}{Untraceable electronic mail, return addresses, and
  digital pseudonyms}.
\newblock \bibinfo{journal}{Communications of the ACM} \bibinfo{volume}{24},
  \bibinfo{pages}{84--90}.
\bibitem[{Cheu et~al.(2019)Cheu, Smith, Ullman, Zeber and
  Zhilyaev}]{37_cheu2019distributed}
\bibinfo{author}{Cheu, A.}, \bibinfo{author}{Smith, A.},
  \bibinfo{author}{Ullman, J.}, \bibinfo{author}{Zeber, D.},
  \bibinfo{author}{Zhilyaev, M.}, \bibinfo{year}{2019}.
\newblock \bibinfo{title}{Distributed differential privacy via shuffling}, in:
  \bibinfo{booktitle}{Annual International Conference on the Theory and
  Applications of Cryptographic Techniques}, \bibinfo{organization}{Springer}.
  pp. \bibinfo{pages}{375--403}.
\bibitem[{Chor et~al.(1994)Chor, Fiat and Naor}]{118_chor1994tracing}
\bibinfo{author}{Chor, B.}, \bibinfo{author}{Fiat, A.}, \bibinfo{author}{Naor,
  M.}, \bibinfo{year}{1994}.
\newblock \bibinfo{title}{Tracing traitors}, in: \bibinfo{booktitle}{Annual
  International Cryptology Conference}, \bibinfo{organization}{Springer}. pp.
  \bibinfo{pages}{257--270}.
\bibitem[{Danezis and Diaz(2008)}]{102_danezis2008survey}
\bibinfo{author}{Danezis, G.}, \bibinfo{author}{Diaz, C.},
  \bibinfo{year}{2008}.
\newblock \bibinfo{title}{A survey of anonymous communication channels}.
\newblock \bibinfo{type}{Technical Report}. Technical Report MSR-TR-2008-35,
  Microsoft Research.
\bibitem[{Delfs and Knebl(2007)}]{116_menezes1996handbook}
\bibinfo{author}{Delfs, H.}, \bibinfo{author}{Knebl, H.}, \bibinfo{year}{2007}.
\newblock \bibinfo{title}{One-Way Functions and the Basic Assumptions. In:
  Introduction to Cryptography. Information Security and Cryptography (Texts
  and Monographs)}.
\newblock \bibinfo{publisher}{Springer}.
\bibitem[{Deng and Liu(2017)}]{205_deng2017consultative}
\bibinfo{author}{Deng, J.}, \bibinfo{author}{Liu, P.}, \bibinfo{year}{2017}.
\newblock \bibinfo{title}{Consultative authoritarianism: the drafting of
  china’s internet security law and e-commerce law}.
\newblock \bibinfo{journal}{Journal of Contemporary China}
  \bibinfo{volume}{26}, \bibinfo{pages}{679--695}.
\bibitem[{Desfontaines and Pej{\'o}(2020)}]{217_desfontaines2020sok}
\bibinfo{author}{Desfontaines, D.}, \bibinfo{author}{Pej{\'o}, B.},
  \bibinfo{year}{2020}.
\newblock \bibinfo{title}{Sok: Differential privacies}.
\newblock \bibinfo{journal}{Proceedings on Privacy Enhancing Technologies}
  \bibinfo{volume}{2020}, \bibinfo{pages}{288--313}.
\bibitem[{Devroye(1986)}]{179_devroye1986sample}
\bibinfo{author}{Devroye, L.}, \bibinfo{year}{1986}.
\newblock \bibinfo{title}{Sample-based non-uniform random variate generation},
  in: \bibinfo{booktitle}{Proceedings of the 18th conference on Winter
  simulation}, pp. \bibinfo{pages}{260--265}.
\bibitem[{Dinur and Nissim(2003)}]{210_dinur2003revealing}
\bibinfo{author}{Dinur, I.}, \bibinfo{author}{Nissim, K.},
  \bibinfo{year}{2003}.
\newblock \bibinfo{title}{Revealing information while preserving privacy}, in:
  \bibinfo{booktitle}{Proceedings of the twenty-second ACM SIGMOD-SIGACT-SIGART
  symposium on Principles of database systems}, pp. \bibinfo{pages}{202--210}.
\bibitem[{Domingo-Ferrer and Soria-Comas(2015)}]{115_domingo2015t}
\bibinfo{author}{Domingo-Ferrer, J.}, \bibinfo{author}{Soria-Comas, J.},
  \bibinfo{year}{2015}.
\newblock \bibinfo{title}{From t-closeness to differential privacy and vice
  versa in data anonymization}.
\newblock \bibinfo{journal}{Knowledge-Based Systems} \bibinfo{volume}{74},
  \bibinfo{pages}{151--158}.
\bibitem[{Dwork(2006)}]{0_dwork2006differential}
\bibinfo{author}{Dwork, C.}, \bibinfo{year}{2006}.
\newblock \bibinfo{title}{Differential privacy in: Proc. of the 33rd
  international colloquium on automata, languages and programming (icalp),
  1--12}.
\bibitem[{Dwork(2008)}]{237_dwork2008differential}
\bibinfo{author}{Dwork, C.}, \bibinfo{year}{2008}.
\newblock \bibinfo{title}{Differential privacy: A survey of results}, in:
  \bibinfo{booktitle}{International conference on theory and applications of
  models of computation}, \bibinfo{organization}{Springer}. pp.
  \bibinfo{pages}{1--19}.
\bibitem[{Dwork(2010)}]{238_dwork2010differential}
\bibinfo{author}{Dwork, C.}, \bibinfo{year}{2010}.
\newblock \bibinfo{title}{Differential privacy in new settings}, in:
  \bibinfo{booktitle}{Proceedings of the twenty-first annual ACM-SIAM symposium
  on Discrete Algorithms}, \bibinfo{organization}{SIAM}. pp.
  \bibinfo{pages}{174--183}.
\bibitem[{Dwork(2011a)}]{239_dwork2011firm}
\bibinfo{author}{Dwork, C.}, \bibinfo{year}{2011}a.
\newblock \bibinfo{title}{A firm foundation for private data analysis}.
\newblock \bibinfo{journal}{Communications of the ACM} \bibinfo{volume}{54},
  \bibinfo{pages}{86--95}.
\bibitem[{Dwork(2011b)}]{224_dwork2011firm}
\bibinfo{author}{Dwork, C.}, \bibinfo{year}{2011}b.
\newblock \bibinfo{title}{A firm foundation for private data analysis}.
\newblock \bibinfo{journal}{Communications of the ACM} \bibinfo{volume}{54},
  \bibinfo{pages}{86--95}.
\bibitem[{Dwork et~al.(2006)Dwork, Kenthapadi, McSherry, Mironov and
  Naor}]{178_dwork2006our}
\bibinfo{author}{Dwork, C.}, \bibinfo{author}{Kenthapadi, K.},
  \bibinfo{author}{McSherry, F.}, \bibinfo{author}{Mironov, I.},
  \bibinfo{author}{Naor, M.}, \bibinfo{year}{2006}.
\newblock \bibinfo{title}{Our data, ourselves: Privacy via distributed noise
  generation}, in: \bibinfo{booktitle}{Annual International Conference on the
  Theory and Applications of Cryptographic Techniques},
  \bibinfo{organization}{Springer}. pp. \bibinfo{pages}{486--503}.
\bibitem[{Dwork et~al.(2016)Dwork, McSherry, Nissim and
  Smith}]{223_dwork2016calibrating}
\bibinfo{author}{Dwork, C.}, \bibinfo{author}{McSherry, F.},
  \bibinfo{author}{Nissim, K.}, \bibinfo{author}{Smith, A.},
  \bibinfo{year}{2016}.
\newblock \bibinfo{title}{Calibrating noise to sensitivity in private data
  analysis}.
\newblock \bibinfo{journal}{Journal of Privacy and Confidentiality}
  \bibinfo{volume}{7}, \bibinfo{pages}{17--51}.
\bibitem[{Dwork et~al.(2010a)Dwork, Naor, Pitassi and
  Rothblum}]{123_dwork2010differential}
\bibinfo{author}{Dwork, C.}, \bibinfo{author}{Naor, M.},
  \bibinfo{author}{Pitassi, T.}, \bibinfo{author}{Rothblum, G.N.},
  \bibinfo{year}{2010}a.
\newblock \bibinfo{title}{Differential privacy under continual observation},
  in: \bibinfo{booktitle}{Proceedings of the forty-second ACM symposium on
  Theory of computing}, pp. \bibinfo{pages}{715--724}.
\bibitem[{Dwork et~al.(2015a)Dwork, Naor, Reingold and
  Rothblum}]{124_dwork2015pure}
\bibinfo{author}{Dwork, C.}, \bibinfo{author}{Naor, M.},
  \bibinfo{author}{Reingold, O.}, \bibinfo{author}{Rothblum, G.N.},
  \bibinfo{year}{2015}a.
\newblock \bibinfo{title}{Pure differential privacy for rectangle queries via
  private partitions}, in: \bibinfo{booktitle}{International Conference on the
  Theory and Application of Cryptology and Information Security},
  \bibinfo{organization}{Springer}. pp. \bibinfo{pages}{735--751}.
\bibitem[{Dwork et~al.(2009)Dwork, Naor, Reingold, Rothblum and
  Vadhan}]{39_10.1145/1536414.1536467}
\bibinfo{author}{Dwork, C.}, \bibinfo{author}{Naor, M.},
  \bibinfo{author}{Reingold, O.}, \bibinfo{author}{Rothblum, G.N.},
  \bibinfo{author}{Vadhan, S.}, \bibinfo{year}{2009}.
\newblock \bibinfo{title}{On the complexity of differentially private data
  release: Efficient algorithms and hardness results},
  \bibinfo{publisher}{Association for Computing Machinery},
  \bibinfo{address}{New York, NY, USA}. p. \bibinfo{pages}{381–390}.
\bibitem[{Dwork et~al.(2015b)Dwork, Nikolov and
  Talwar}]{129_dwork2015efficient}
\bibinfo{author}{Dwork, C.}, \bibinfo{author}{Nikolov, A.},
  \bibinfo{author}{Talwar, K.}, \bibinfo{year}{2015}b.
\newblock \bibinfo{title}{Efficient algorithms for privately releasing
  marginals via convex relaxations}.
\newblock \bibinfo{journal}{Discrete \& Computational Geometry}
  \bibinfo{volume}{53}, \bibinfo{pages}{650--673}.
\bibitem[{Dwork and Nissim(2004)}]{211_dwork2004privacy}
\bibinfo{author}{Dwork, C.}, \bibinfo{author}{Nissim, K.},
  \bibinfo{year}{2004}.
\newblock \bibinfo{title}{Privacy-preserving datamining on vertically
  partitioned databases}, in: \bibinfo{booktitle}{Annual International
  Cryptology Conference}, \bibinfo{organization}{Springer}. pp.
  \bibinfo{pages}{528--544}.
\bibitem[{Dwork et~al.(2014)Dwork, Roth et~al.}]{220_dwork2014algorithmic}
\bibinfo{author}{Dwork, C.}, \bibinfo{author}{Roth, A.}, et~al.,
  \bibinfo{year}{2014}.
\newblock \bibinfo{title}{The algorithmic foundations of differential privacy.}
\newblock \bibinfo{journal}{Foundations and Trends in Theoretical Computer
  Science} \bibinfo{volume}{9}, \bibinfo{pages}{211--407}.
\bibitem[{Dwork et~al.(2010b)Dwork, Rothblum and
  Vadhan}]{133_dwork2010boosting}
\bibinfo{author}{Dwork, C.}, \bibinfo{author}{Rothblum, G.N.},
  \bibinfo{author}{Vadhan, S.}, \bibinfo{year}{2010}b.
\newblock \bibinfo{title}{Boosting and differential privacy}, in:
  \bibinfo{booktitle}{2010 IEEE 51st Annual Symposium on Foundations of
  Computer Science}, \bibinfo{organization}{IEEE}. pp. \bibinfo{pages}{51--60}.
\bibitem[{Eigner et~al.(2015)Eigner, Kate, Maffei and
  Pampaloni}]{192_eigner2015achieving}
\bibinfo{author}{Eigner, F.}, \bibinfo{author}{Kate, A.},
  \bibinfo{author}{Maffei, M.}, \bibinfo{author}{Pampaloni, F.},
  \bibinfo{year}{2015}.
\newblock \bibinfo{title}{Achieving optimal utility for distributed
  differential privacy using secure multiparty computation}.
\newblock \bibinfo{journal}{Applications of Secure Multiparty Computation}
  \bibinfo{volume}{13}.
\bibitem[{ElGamal(1985)}]{155_elgamal1985public}
\bibinfo{author}{ElGamal, T.}, \bibinfo{year}{1985}.
\newblock \bibinfo{title}{A public key cryptosystem and a signature scheme
  based on discrete logarithms}.
\newblock \bibinfo{journal}{IEEE transactions on information theory}
  \bibinfo{volume}{31}, \bibinfo{pages}{469--472}.
\bibitem[{Erlingsson et~al.(2019)Erlingsson, Feldman, Mironov, Raghunathan,
  Talwar and Thakurta}]{35_erlingsson2019amplification}
\bibinfo{author}{Erlingsson, {\'U}.}, \bibinfo{author}{Feldman, V.},
  \bibinfo{author}{Mironov, I.}, \bibinfo{author}{Raghunathan, A.},
  \bibinfo{author}{Talwar, K.}, \bibinfo{author}{Thakurta, A.},
  \bibinfo{year}{2019}.
\newblock \bibinfo{title}{Amplification by shuffling: From local to central
  differential privacy via anonymity}, in: \bibinfo{booktitle}{Proceedings of
  the Thirtieth Annual ACM-SIAM Symposium on Discrete Algorithms},
  \bibinfo{organization}{SIAM}. pp. \bibinfo{pages}{2468--2479}.
\bibitem[{Erlingsson et~al.(2014)Erlingsson, Pihur and
  Korolova}]{207_erlingsson2014rappor}
\bibinfo{author}{Erlingsson, {\'U}.}, \bibinfo{author}{Pihur, V.},
  \bibinfo{author}{Korolova, A.}, \bibinfo{year}{2014}.
\newblock \bibinfo{title}{Rappor: Randomized aggregatable privacy-preserving
  ordinal response}, in: \bibinfo{booktitle}{Proceedings of the 2014 ACM SIGSAC
  conference on computer and communications security}, pp.
  \bibinfo{pages}{1054--1067}.
\bibitem[{Fanti et~al.(2016)Fanti, Pihur and
  Erlingsson}]{106_fanti2016building}
\bibinfo{author}{Fanti, G.}, \bibinfo{author}{Pihur, V.},
  \bibinfo{author}{Erlingsson, {\'U}.}, \bibinfo{year}{2016}.
\newblock \bibinfo{title}{Building a rappor with the unknown:
  Privacy-preserving learning of associations and data dictionaries}.
\newblock \bibinfo{journal}{Proceedings on Privacy Enhancing Technologies}
  \bibinfo{volume}{2016}, \bibinfo{pages}{41--61}.
\bibitem[{Feldman et~al.(2018)Feldman, Mironov, Talwar and
  Thakurta}]{230_feldman2018privacy}
\bibinfo{author}{Feldman, V.}, \bibinfo{author}{Mironov, I.},
  \bibinfo{author}{Talwar, K.}, \bibinfo{author}{Thakurta, A.},
  \bibinfo{year}{2018}.
\newblock \bibinfo{title}{Privacy amplification by iteration}, in:
  \bibinfo{booktitle}{2018 IEEE 59th Annual Symposium on Foundations of
  Computer Science (FOCS)}, \bibinfo{organization}{IEEE}. pp.
  \bibinfo{pages}{521--532}.
\bibitem[{Fung et~al.(2005)Fung, Wang and Yu}]{111_fung2005top}
\bibinfo{author}{Fung, B.C.}, \bibinfo{author}{Wang, K.}, \bibinfo{author}{Yu,
  P.S.}, \bibinfo{year}{2005}.
\newblock \bibinfo{title}{Top-down specialization for information and privacy
  preservation}, in: \bibinfo{booktitle}{21st international conference on data
  engineering (ICDE'05)}, \bibinfo{organization}{IEEE}. pp.
  \bibinfo{pages}{205--216}.
\bibitem[{Gabizon et~al.(2006)Gabizon, Raz and
  Shaltiel}]{177_gabizon2006deterministic}
\bibinfo{author}{Gabizon, A.}, \bibinfo{author}{Raz, R.},
  \bibinfo{author}{Shaltiel, R.}, \bibinfo{year}{2006}.
\newblock \bibinfo{title}{Deterministic extractors for bit-fixing sources by
  obtaining an independent seed}.
\newblock \bibinfo{journal}{SIAM Journal on Computing} \bibinfo{volume}{36},
  \bibinfo{pages}{1072--1094}.
\bibitem[{Gentry and Boneh(2009)}]{152_gentry2009fully}
\bibinfo{author}{Gentry, C.}, \bibinfo{author}{Boneh, D.},
  \bibinfo{year}{2009}.
\newblock \bibinfo{title}{A fully homomorphic encryption scheme}.
  volume~\bibinfo{volume}{20}.
\newblock \bibinfo{publisher}{Stanford university Stanford}.
\bibitem[{Ghosh et~al.(2012)Ghosh, Roughgarden and
  Sundararajan}]{173_ghosh2012universally}
\bibinfo{author}{Ghosh, A.}, \bibinfo{author}{Roughgarden, T.},
  \bibinfo{author}{Sundararajan, M.}, \bibinfo{year}{2012}.
\newblock \bibinfo{title}{Universally utility-maximizing privacy mechanisms}.
\newblock \bibinfo{journal}{SIAM Journal on Computing} \bibinfo{volume}{41},
  \bibinfo{pages}{1673--1693}.
\bibitem[{Goldwasser and Micali(1982)}]{154_10.1145/800070.802212}
\bibinfo{author}{Goldwasser, S.}, \bibinfo{author}{Micali, S.},
  \bibinfo{year}{1982}.
\newblock \bibinfo{title}{Probabilistic encryption \& how to play mental poker
  keeping secret all partial information}, in: \bibinfo{booktitle}{Proceedings
  of the Fourteenth Annual ACM Symposium on Theory of Computing},
  \bibinfo{publisher}{Association for Computing Machinery},
  \bibinfo{address}{New York, NY, USA}. p. \bibinfo{pages}{365–377}.
\bibitem[{Goldwasser and Micali(1984)}]{213_goldwasser1984probabilistic}
\bibinfo{author}{Goldwasser, S.}, \bibinfo{author}{Micali, S.},
  \bibinfo{year}{1984}.
\newblock \bibinfo{title}{Probabilistic encryption}.
\newblock \bibinfo{journal}{Journal of computer and system sciences}
  \bibinfo{volume}{28}, \bibinfo{pages}{270--299}.
\bibitem[{Goryczka and Xiong(2015)}]{45_goryczka2015comprehensive}
\bibinfo{author}{Goryczka, S.}, \bibinfo{author}{Xiong, L.},
  \bibinfo{year}{2015}.
\newblock \bibinfo{title}{A comprehensive comparison of multiparty secure
  additions with differential privacy}.
\newblock \bibinfo{journal}{IEEE transactions on dependable and secure
  computing} \bibinfo{volume}{14}, \bibinfo{pages}{463--477}.
\bibitem[{Goryczka et~al.(2013)Goryczka, Xiong and
  Sunderam}]{166_goryczka2013secure}
\bibinfo{author}{Goryczka, S.}, \bibinfo{author}{Xiong, L.},
  \bibinfo{author}{Sunderam, V.}, \bibinfo{year}{2013}.
\newblock \bibinfo{title}{Secure multiparty aggregation with differential
  privacy: A comparative study}, in: \bibinfo{booktitle}{Proceedings of the
  Joint EDBT/ICDT 2013 Workshops}, pp. \bibinfo{pages}{155--163}.
\bibitem[{Groce et~al.(2019)Groce, Rindal and Rosulek}]{182_groce2019cheaper}
\bibinfo{author}{Groce, A.}, \bibinfo{author}{Rindal, P.},
  \bibinfo{author}{Rosulek, M.}, \bibinfo{year}{2019}.
\newblock \bibinfo{title}{Cheaper private set intersection via differentially
  private leakage}.
\newblock \bibinfo{journal}{Proceedings on Privacy Enhancing Technologies}
  \bibinfo{volume}{2019}, \bibinfo{pages}{6--25}.
\bibitem[{Gupta et~al.(2013)Gupta, Hardt, Roth and
  Ullman}]{137_gupta2013privately}
\bibinfo{author}{Gupta, A.}, \bibinfo{author}{Hardt, M.},
  \bibinfo{author}{Roth, A.}, \bibinfo{author}{Ullman, J.},
  \bibinfo{year}{2013}.
\newblock \bibinfo{title}{Privately releasing conjunctions and the statistical
  query barrier}.
\newblock \bibinfo{journal}{SIAM Journal on Computing} \bibinfo{volume}{42},
  \bibinfo{pages}{1494--1520}.
\bibitem[{Gupta et~al.(2012)Gupta, Roth and Ullman}]{138_gupta2012iterative}
\bibinfo{author}{Gupta, A.}, \bibinfo{author}{Roth, A.},
  \bibinfo{author}{Ullman, J.}, \bibinfo{year}{2012}.
\newblock \bibinfo{title}{Iterative constructions and private data release},
  in: \bibinfo{booktitle}{Theory of cryptography conference},
  \bibinfo{organization}{Springer}. pp. \bibinfo{pages}{339--356}.
\bibitem[{Gursoy et~al.(2019)Gursoy, Tamersoy, Truex, Wei and
  Liu}]{105_gursoy2019secure}
\bibinfo{author}{Gursoy, M.E.}, \bibinfo{author}{Tamersoy, A.},
  \bibinfo{author}{Truex, S.}, \bibinfo{author}{Wei, W.}, \bibinfo{author}{Liu,
  L.}, \bibinfo{year}{2019}.
\newblock \bibinfo{title}{Secure and utility-aware data collection with
  condensed local differential privacy}.
\newblock \bibinfo{journal}{IEEE Transactions on Dependable and Secure
  Computing} .
\bibitem[{Harding et~al.(2019)Harding, Vanto, Clark, Hannah~Ji and
  Ainsworth}]{204}
\bibinfo{author}{Harding, E.L.}, \bibinfo{author}{Vanto, J.J.},
  \bibinfo{author}{Clark, R.}, \bibinfo{author}{Hannah~Ji, L.},
  \bibinfo{author}{Ainsworth, S.C.}, \bibinfo{year}{2019}.
\newblock \bibinfo{title}{Understanding the scope and impact of the california
  consumer privacy act of 2018}.
\newblock \bibinfo{journal}{Journal of Data Protection \& Privacy}
  \bibinfo{volume}{2}, \bibinfo{pages}{234--253}.
\bibitem[{Hardt et~al.(2012a)Hardt, Ligett and McSherry}]{135_hardt2012simple}
\bibinfo{author}{Hardt, M.}, \bibinfo{author}{Ligett, K.},
  \bibinfo{author}{McSherry, F.}, \bibinfo{year}{2012}a.
\newblock \bibinfo{title}{A simple and practical algorithm for differentially
  private data release}, in: \bibinfo{booktitle}{Advances in Neural Information
  Processing Systems}, pp. \bibinfo{pages}{2339--2347}.
\bibitem[{Hardt et~al.(2012b)Hardt, Rothblum and
  Servedio}]{139_hardt2012private}
\bibinfo{author}{Hardt, M.}, \bibinfo{author}{Rothblum, G.N.},
  \bibinfo{author}{Servedio, R.A.}, \bibinfo{year}{2012}b.
\newblock \bibinfo{title}{Private data release via learning thresholds}, in:
  \bibinfo{booktitle}{Proceedings of the twenty-third annual ACM-SIAM symposium
  on Discrete Algorithms}, \bibinfo{organization}{SIAM}. pp.
  \bibinfo{pages}{168--187}.
\bibitem[{Hassan et~al.(2019)Hassan, Rehmani and
  Chen}]{216_hassan2019differential}
\bibinfo{author}{Hassan, M.U.}, \bibinfo{author}{Rehmani, M.H.},
  \bibinfo{author}{Chen, J.}, \bibinfo{year}{2019}.
\newblock \bibinfo{title}{Differential privacy techniques for cyber physical
  systems: a survey}.
\newblock \bibinfo{journal}{IEEE Communications Surveys \& Tutorials}
  \bibinfo{volume}{22}, \bibinfo{pages}{746--789}.
\bibitem[{Hayes and Ohrimenko(2018)}]{108_hayes2018contamination}
\bibinfo{author}{Hayes, J.}, \bibinfo{author}{Ohrimenko, O.},
  \bibinfo{year}{2018}.
\newblock \bibinfo{title}{Contamination attacks and mitigation in multi-party
  machine learning}, in: \bibinfo{booktitle}{Advances in Neural Information
  Processing Systems}, pp. \bibinfo{pages}{6604--6615}.
\bibitem[{Huang et~al.(2019a)Huang, Liao, Zhou and
  Chen}]{117_huang2019efficient}
\bibinfo{author}{Huang, W.}, \bibinfo{author}{Liao, Y.}, \bibinfo{author}{Zhou,
  S.}, \bibinfo{author}{Chen, H.}, \bibinfo{year}{2019}a.
\newblock \bibinfo{title}{An efficient deniable authenticated encryption scheme
  for privacy protection}.
\newblock \bibinfo{journal}{IEEE Access} \bibinfo{volume}{7},
  \bibinfo{pages}{43453--43461}.
\bibitem[{Huang et~al.(2019b)Huang, Zhou, Liao and
  Chen}]{150_huang2019efficient}
\bibinfo{author}{Huang, W.}, \bibinfo{author}{Zhou, S.}, \bibinfo{author}{Liao,
  Y.}, \bibinfo{author}{Chen, H.}, \bibinfo{year}{2019}b.
\newblock \bibinfo{title}{An efficient differential privacy logistic
  classification mechanism}.
\newblock \bibinfo{journal}{IEEE Internet of Things Journal}
  \bibinfo{volume}{6}, \bibinfo{pages}{10620--10626}.
\bibitem[{Ishai and Paskin(2007)}]{159_ishai2007evaluating}
\bibinfo{author}{Ishai, Y.}, \bibinfo{author}{Paskin, A.},
  \bibinfo{year}{2007}.
\newblock \bibinfo{title}{Evaluating branching programs on encrypted data}, in:
  \bibinfo{booktitle}{Theory of Cryptography Conference},
  \bibinfo{organization}{Springer}. pp. \bibinfo{pages}{575--594}.
\bibitem[{Ji et~al.(2014)Ji, Lipton and Elkan}]{235_ji2014differential}
\bibinfo{author}{Ji, Z.}, \bibinfo{author}{Lipton, Z.C.},
  \bibinfo{author}{Elkan, C.}, \bibinfo{year}{2014}.
\newblock \bibinfo{title}{Differential privacy and machine learning: a survey
  and review}.
\newblock \bibinfo{journal}{arXiv preprint arXiv:1412.7584} .
\bibitem[{Johnson et~al.(2018)Johnson, Near and Song}]{208_johnson2018towards}
\bibinfo{author}{Johnson, N.}, \bibinfo{author}{Near, J.P.},
  \bibinfo{author}{Song, D.}, \bibinfo{year}{2018}.
\newblock \bibinfo{title}{Towards practical differential privacy for sql
  queries}.
\newblock \bibinfo{journal}{Proceedings of the VLDB Endowment}
  \bibinfo{volume}{11}, \bibinfo{pages}{526--539}.
\bibitem[{Johnson et~al.(2005)Johnson, Kemp and
  Kotz}]{174_johnson2005univariate}
\bibinfo{author}{Johnson, N.L.}, \bibinfo{author}{Kemp, A.W.},
  \bibinfo{author}{Kotz, S.}, \bibinfo{year}{2005}.
\newblock \bibinfo{title}{Univariate discrete distributions}. volume
  \bibinfo{volume}{444}.
\newblock \bibinfo{publisher}{John Wiley \& Sons}.
\bibitem[{Kacsmar et~al.()Kacsmar, Khurram, Lukas, Norton, Shafieinejad, Shang,
  Baseri, Sepehri, Oya and Kerschbaum}]{181_kacsmardifferentially}
\bibinfo{author}{Kacsmar, B.}, \bibinfo{author}{Khurram, B.},
  \bibinfo{author}{Lukas, N.}, \bibinfo{author}{Norton, A.},
  \bibinfo{author}{Shafieinejad, M.}, \bibinfo{author}{Shang, Z.},
  \bibinfo{author}{Baseri, Y.}, \bibinfo{author}{Sepehri, M.},
  \bibinfo{author}{Oya, S.}, \bibinfo{author}{Kerschbaum, F.}, .
\newblock \bibinfo{title}{Differentially private two-party set operations} .
\bibitem[{Kairouz et~al.(2015)Kairouz, Oh and
  Viswanath}]{145_kairouz2015composition}
\bibinfo{author}{Kairouz, P.}, \bibinfo{author}{Oh, S.},
  \bibinfo{author}{Viswanath, P.}, \bibinfo{year}{2015}.
\newblock \bibinfo{title}{The composition theorem for differential privacy},
  in: \bibinfo{booktitle}{International conference on machine learning},
  \bibinfo{organization}{PMLR}. pp. \bibinfo{pages}{1376--1385}.
\bibitem[{Kasiviswanathan et~al.(2011)Kasiviswanathan, Lee, Nissim,
  Raskhodnikova and Smith}]{110_kasiviswanathan2011can}
\bibinfo{author}{Kasiviswanathan, S.P.}, \bibinfo{author}{Lee, H.K.},
  \bibinfo{author}{Nissim, K.}, \bibinfo{author}{Raskhodnikova, S.},
  \bibinfo{author}{Smith, A.}, \bibinfo{year}{2011}.
\newblock \bibinfo{title}{What can we learn privately?}
\newblock \bibinfo{journal}{SIAM Journal on Computing} \bibinfo{volume}{40},
  \bibinfo{pages}{793--826}.
\bibitem[{Kasiviswanathan et~al.(2010)Kasiviswanathan, Rudelson, Smith and
  Ullman}]{142_kasiviswanathan2010price}
\bibinfo{author}{Kasiviswanathan, S.P.}, \bibinfo{author}{Rudelson, M.},
  \bibinfo{author}{Smith, A.}, \bibinfo{author}{Ullman, J.},
  \bibinfo{year}{2010}.
\newblock \bibinfo{title}{The price of privately releasing contingency tables
  and the spectra of random matrices with correlated rows}, in:
  \bibinfo{booktitle}{Proceedings of the forty-second ACM symposium on Theory
  of computing}, pp. \bibinfo{pages}{775--784}.
\bibitem[{Kearns(1998)}]{119_10.1145/293347.293351}
\bibinfo{author}{Kearns, M.}, \bibinfo{year}{1998}.
\newblock \bibinfo{title}{Efficient noise-tolerant learning from statistical
  queries}.
\newblock \bibinfo{journal}{J. ACM} \bibinfo{volume}{45},
  \bibinfo{pages}{983–1006}.
\bibitem[{Kiayias and Yung(2001)}]{121_kiayias2001crafty}
\bibinfo{author}{Kiayias, A.}, \bibinfo{author}{Yung, M.},
  \bibinfo{year}{2001}.
\newblock \bibinfo{title}{On crafty pirates and foxy tracers}, in:
  \bibinfo{booktitle}{ACM Workshop on Digital Rights Management},
  \bibinfo{organization}{Springer}. pp. \bibinfo{pages}{22--39}.
\bibitem[{Kim et~al.(2019)Kim, Lee, Ohno-Machado and Jiang}]{46_kim2019secure}
\bibinfo{author}{Kim, M.}, \bibinfo{author}{Lee, J.},
  \bibinfo{author}{Ohno-Machado, L.}, \bibinfo{author}{Jiang, X.},
  \bibinfo{year}{2019}.
\newblock \bibinfo{title}{Secure and differentially private logistic regression
  for horizontally distributed data}.
\newblock \bibinfo{journal}{IEEE Transactions on Information Forensics and
  Security} \bibinfo{volume}{15}, \bibinfo{pages}{695--710}.
\bibitem[{King(2011)}]{201_king2011ensuring}
\bibinfo{author}{King, G.}, \bibinfo{year}{2011}.
\newblock \bibinfo{title}{Ensuring the data-rich future of the social
  sciences}.
\newblock \bibinfo{journal}{science} \bibinfo{volume}{331},
  \bibinfo{pages}{719--721}.
\bibitem[{Kisilevich et~al.(2009)Kisilevich, Rokach, Elovici and
  Shapira}]{112_kisilevich2009efficient}
\bibinfo{author}{Kisilevich, S.}, \bibinfo{author}{Rokach, L.},
  \bibinfo{author}{Elovici, Y.}, \bibinfo{author}{Shapira, B.},
  \bibinfo{year}{2009}.
\newblock \bibinfo{title}{Efficient multidimensional suppression for
  k-anonymity}.
\newblock \bibinfo{journal}{IEEE Transactions on Knowledge and Data
  Engineering} \bibinfo{volume}{22}, \bibinfo{pages}{334--347}.
\bibitem[{Kotz et~al.(2012)Kotz, Kozubowski and
  Podgorski}]{163_kotz2012laplace}
\bibinfo{author}{Kotz, S.}, \bibinfo{author}{Kozubowski, T.},
  \bibinfo{author}{Podgorski, K.}, \bibinfo{year}{2012}.
\newblock \bibinfo{title}{The Laplace distribution and generalizations: a
  revisit with applications to communications, economics, engineering, and
  finance}.
\newblock \bibinfo{publisher}{Springer Science \& Business Media}.
\bibitem[{Kowalczyk et~al.(2018)Kowalczyk, Malkin, Ullman and
  Wichs}]{13_kowalczyk2018hardness}
\bibinfo{author}{Kowalczyk, L.}, \bibinfo{author}{Malkin, T.},
  \bibinfo{author}{Ullman, J.}, \bibinfo{author}{Wichs, D.},
  \bibinfo{year}{2018}.
\newblock \bibinfo{title}{Hardness of non-interactive differential privacy from
  one-way functions}, in: \bibinfo{booktitle}{Annual International Cryptology
  Conference}, \bibinfo{organization}{Springer}. pp. \bibinfo{pages}{437--466}.
\bibitem[{Lang(2011)}]{200_lang2011advancing}
\bibinfo{author}{Lang, T.}, \bibinfo{year}{2011}.
\newblock \bibinfo{title}{Advancing global health research through digital
  technology and sharing data}.
\newblock \bibinfo{journal}{Science} \bibinfo{volume}{331},
  \bibinfo{pages}{714--717}.
\bibitem[{Li et~al.(2020)Li, Ye, Li, Wang, Lou, Hou, Liu and
  Lu}]{195_li2020efficient}
\bibinfo{author}{Li, J.}, \bibinfo{author}{Ye, H.}, \bibinfo{author}{Li, T.},
  \bibinfo{author}{Wang, W.}, \bibinfo{author}{Lou, W.}, \bibinfo{author}{Hou,
  T.}, \bibinfo{author}{Liu, J.}, \bibinfo{author}{Lu, R.},
  \bibinfo{year}{2020}.
\newblock \bibinfo{title}{Efficient and secure outsourcing of differentially
  private data publishing with multiple evaluators}.
\newblock \bibinfo{journal}{IEEE Transactions on Dependable and Secure
  Computing} .
\bibitem[{Li et~al.(2007)Li, Li and Venkatasubramanian}]{113_li2007t}
\bibinfo{author}{Li, N.}, \bibinfo{author}{Li, T.},
  \bibinfo{author}{Venkatasubramanian, S.}, \bibinfo{year}{2007}.
\newblock \bibinfo{title}{t-closeness: Privacy beyond k-anonymity and
  l-diversity}, in: \bibinfo{booktitle}{2007 IEEE 23rd International Conference
  on Data Engineering}, \bibinfo{organization}{IEEE}. pp.
  \bibinfo{pages}{106--115}.
\bibitem[{Li et~al.(2011)Li, Qardaji and Su}]{38_li2011provably}
\bibinfo{author}{Li, N.}, \bibinfo{author}{Qardaji, W.H.}, \bibinfo{author}{Su,
  D.}, \bibinfo{year}{2011}.
\newblock \bibinfo{title}{Provably private data anonymization: Or, k-anonymity
  meets differential privacy}.
\newblock \bibinfo{journal}{CoRR, abs/1101.2604} \bibinfo{volume}{49},
  \bibinfo{pages}{55}.
\bibitem[{L{\'o}pez-Alt et~al.(2012)L{\'o}pez-Alt, Tromer and
  Vaikuntanathan}]{162_lopez2012fly}
\bibinfo{author}{L{\'o}pez-Alt, A.}, \bibinfo{author}{Tromer, E.},
  \bibinfo{author}{Vaikuntanathan, V.}, \bibinfo{year}{2012}.
\newblock \bibinfo{title}{On-the-fly multiparty computation on the cloud via
  multikey fully homomorphic encryption}, in: \bibinfo{booktitle}{Proceedings
  of the forty-fourth annual ACM symposium on Theory of computing}, pp.
  \bibinfo{pages}{1219--1234}.
\bibitem[{Machanavajjhala et~al.(2008)Machanavajjhala, Kifer, Abowd, Gehrke and
  Vilhuber}]{209_machanavajjhala2008privacy}
\bibinfo{author}{Machanavajjhala, A.}, \bibinfo{author}{Kifer, D.},
  \bibinfo{author}{Abowd, J.}, \bibinfo{author}{Gehrke, J.},
  \bibinfo{author}{Vilhuber, L.}, \bibinfo{year}{2008}.
\newblock \bibinfo{title}{Privacy: Theory meets practice on the map}, in:
  \bibinfo{booktitle}{2008 IEEE 24th international conference on data
  engineering}, \bibinfo{organization}{IEEE}. pp. \bibinfo{pages}{277--286}.
\bibitem[{Machanavajjhala et~al.(2007)Machanavajjhala, Kifer, Gehrke and
  Venkitasubramaniam}]{114_machanavajjhala2007diversity}
\bibinfo{author}{Machanavajjhala, A.}, \bibinfo{author}{Kifer, D.},
  \bibinfo{author}{Gehrke, J.}, \bibinfo{author}{Venkitasubramaniam, M.},
  \bibinfo{year}{2007}.
\newblock \bibinfo{title}{l-diversity: Privacy beyond k-anonymity}.
\newblock \bibinfo{journal}{ACM Transactions on Knowledge Discovery from Data
  (TKDD)} \bibinfo{volume}{1}, \bibinfo{pages}{3--es}.
\bibitem[{Marsaglia and Bray(1964)}]{170_marsaglia1964convenient}
\bibinfo{author}{Marsaglia, G.}, \bibinfo{author}{Bray, T.A.},
  \bibinfo{year}{1964}.
\newblock \bibinfo{title}{A convenient method for generating normal variables}.
\newblock \bibinfo{journal}{SIAM review} \bibinfo{volume}{6},
  \bibinfo{pages}{260--264}.
\bibitem[{Marsaglia et~al.(2000)Marsaglia, Tsang
  et~al.}]{171_marsaglia2000ziggurat}
\bibinfo{author}{Marsaglia, G.}, \bibinfo{author}{Tsang, W.W.}, et~al.,
  \bibinfo{year}{2000}.
\newblock \bibinfo{title}{The ziggurat method for generating random variables}.
\newblock \bibinfo{journal}{Journal of statistical software}
  \bibinfo{volume}{5}, \bibinfo{pages}{1--7}.
\bibitem[{Marx(2013)}]{199_marx2013biology}
\bibinfo{author}{Marx, V.}, \bibinfo{year}{2013}.
\newblock \bibinfo{title}{Biology: The big challenges of big data}.
\bibitem[{Matou{\v{s}}ek et~al.(2020)Matou{\v{s}}ek, Nikolov and
  Talwar}]{143_matouvsek2020factorization}
\bibinfo{author}{Matou{\v{s}}ek, J.}, \bibinfo{author}{Nikolov, A.},
  \bibinfo{author}{Talwar, K.}, \bibinfo{year}{2020}.
\newblock \bibinfo{title}{Factorization norms and hereditary discrepancy}.
\newblock \bibinfo{journal}{International Mathematics Research Notices}
  \bibinfo{volume}{2020}, \bibinfo{pages}{751--780}.
\bibitem[{Maurer et~al.(2007)Maurer, Abadi, Anderson, Bellare, Goldreich,
  Okamoto, van Oorschot, Pfitzmann and Rubin}]{146_maurer2007information}
\bibinfo{author}{Maurer, U.}, \bibinfo{author}{Abadi, M.},
  \bibinfo{author}{Anderson, R.}, \bibinfo{author}{Bellare, M.},
  \bibinfo{author}{Goldreich, O.}, \bibinfo{author}{Okamoto, T.},
  \bibinfo{author}{van Oorschot, P.}, \bibinfo{author}{Pfitzmann, B.},
  \bibinfo{author}{Rubin, A.D.}, \bibinfo{year}{2007}.
\newblock \bibinfo{title}{Information security and cryptography}.
\bibitem[{Mazloom and Gordon(2018)}]{196_mazloom2018secure}
\bibinfo{author}{Mazloom, S.}, \bibinfo{author}{Gordon, S.D.},
  \bibinfo{year}{2018}.
\newblock \bibinfo{title}{Secure computation with differentially private access
  patterns}, in: \bibinfo{booktitle}{Proceedings of the 2018 ACM SIGSAC
  Conference on Computer and Communications Security}, pp.
  \bibinfo{pages}{490--507}.
\bibitem[{McSherry and Talwar(2007)}]{222_mcsherry2007mechanism}
\bibinfo{author}{McSherry, F.}, \bibinfo{author}{Talwar, K.},
  \bibinfo{year}{2007}.
\newblock \bibinfo{title}{Mechanism design via differential privacy}, in:
  \bibinfo{booktitle}{48th Annual IEEE Symposium on Foundations of Computer
  Science (FOCS'07)}, \bibinfo{organization}{IEEE}. pp.
  \bibinfo{pages}{94--103}.
\bibitem[{Murakami and Kawamoto(2019)}]{234_murakami2019utility}
\bibinfo{author}{Murakami, T.}, \bibinfo{author}{Kawamoto, Y.},
  \bibinfo{year}{2019}.
\newblock \bibinfo{title}{Utility-optimized local differential privacy
  mechanisms for distribution estimation}, in: \bibinfo{booktitle}{28th
  $\{$USENIX$\}$ Security Symposium ($\{$USENIX$\}$ Security 19)}, pp.
  \bibinfo{pages}{1877--1894}.
\bibitem[{Muthukrishnan and Nikolov(2012)}]{141_muthukrishnan2012optimal}
\bibinfo{author}{Muthukrishnan, S.}, \bibinfo{author}{Nikolov, A.},
  \bibinfo{year}{2012}.
\newblock \bibinfo{title}{Optimal private halfspace counting via discrepancy},
  in: \bibinfo{booktitle}{Proceedings of the forty-fourth annual ACM symposium
  on Theory of computing}, pp. \bibinfo{pages}{1285--1292}.
\bibitem[{Nikolov et~al.(2013)Nikolov, Talwar and
  Zhang}]{140_nikolov2013geometry}
\bibinfo{author}{Nikolov, A.}, \bibinfo{author}{Talwar, K.},
  \bibinfo{author}{Zhang, L.}, \bibinfo{year}{2013}.
\newblock \bibinfo{title}{The geometry of differential privacy: the sparse and
  approximate cases}, in: \bibinfo{booktitle}{Proceedings of the forty-fifth
  annual ACM symposium on Theory of computing}, pp. \bibinfo{pages}{351--360}.
\bibitem[{Nissim et~al.(2007)Nissim, Raskhodnikova and
  Smith}]{189_nissim2007smooth}
\bibinfo{author}{Nissim, K.}, \bibinfo{author}{Raskhodnikova, S.},
  \bibinfo{author}{Smith, A.}, \bibinfo{year}{2007}.
\newblock \bibinfo{title}{Smooth sensitivity and sampling in private data
  analysis}, in: \bibinfo{booktitle}{Proceedings of the thirty-ninth annual ACM
  symposium on Theory of computing}, pp. \bibinfo{pages}{75--84}.
\bibitem[{Overpeck et~al.(2011)Overpeck, Meehl, Bony and
  Easterling}]{198_overpeck2011climate}
\bibinfo{author}{Overpeck, J.T.}, \bibinfo{author}{Meehl, G.A.},
  \bibinfo{author}{Bony, S.}, \bibinfo{author}{Easterling, D.R.},
  \bibinfo{year}{2011}.
\newblock \bibinfo{title}{Climate data challenges in the 21st century}.
\newblock \bibinfo{journal}{science} \bibinfo{volume}{331},
  \bibinfo{pages}{700--702}.
\bibitem[{Paillier(1999)}]{157_paillier1999public}
\bibinfo{author}{Paillier, P.}, \bibinfo{year}{1999}.
\newblock \bibinfo{title}{Public-key cryptosystems based on composite degree
  residuosity classes}, in: \bibinfo{booktitle}{International conference on the
  theory and applications of cryptographic techniques},
  \bibinfo{organization}{Springer}. pp. \bibinfo{pages}{223--238}.
\bibitem[{Pinkas et~al.(2015)Pinkas, Schneider, Segev and
  Zohner}]{183_pinkas2015phasing}
\bibinfo{author}{Pinkas, B.}, \bibinfo{author}{Schneider, T.},
  \bibinfo{author}{Segev, G.}, \bibinfo{author}{Zohner, M.},
  \bibinfo{year}{2015}.
\newblock \bibinfo{title}{Phasing: Private set intersection using
  permutation-based hashing}, in: \bibinfo{booktitle}{24th $\{$USENIX$\}$
  Security Symposium ($\{$USENIX$\}$ Security 15)}, pp.
  \bibinfo{pages}{515--530}.
\bibitem[{Pinkas et~al.(2018)Pinkas, Schneider and
  Zohner}]{184_pinkas2018scalable}
\bibinfo{author}{Pinkas, B.}, \bibinfo{author}{Schneider, T.},
  \bibinfo{author}{Zohner, M.}, \bibinfo{year}{2018}.
\newblock \bibinfo{title}{Scalable private set intersection based on ot
  extension}.
\newblock \bibinfo{journal}{ACM Transactions on Privacy and Security (TOPS)}
  \bibinfo{volume}{21}, \bibinfo{pages}{1--35}.
\bibitem[{Rastogi and Nath(2010)}]{165_rastogi2010differentially}
\bibinfo{author}{Rastogi, V.}, \bibinfo{author}{Nath, S.},
  \bibinfo{year}{2010}.
\newblock \bibinfo{title}{Differentially private aggregation of distributed
  time-series with transformation and encryption}, in:
  \bibinfo{booktitle}{Proceedings of the 2010 ACM SIGMOD International
  Conference on Management of data}, pp. \bibinfo{pages}{735--746}.
\bibitem[{Regulation(2018)}]{203}
\bibinfo{author}{Regulation, P.}, \bibinfo{year}{2018}.
\newblock \bibinfo{title}{General data protection regulation}.
\newblock \bibinfo{journal}{Intouch} .
\bibitem[{Rivest et~al.(1978a)Rivest, Adleman, Dertouzos
  et~al.}]{151_rivest1978data}
\bibinfo{author}{Rivest, R.L.}, \bibinfo{author}{Adleman, L.},
  \bibinfo{author}{Dertouzos, M.L.}, et~al., \bibinfo{year}{1978}a.
\newblock \bibinfo{title}{On data banks and privacy homomorphisms}.
\newblock \bibinfo{journal}{Foundations of secure computation}
  \bibinfo{volume}{4}, \bibinfo{pages}{169--180}.
\bibitem[{Rivest et~al.(1978b)Rivest, Shamir and
  Adleman}]{153_rivest1978method}
\bibinfo{author}{Rivest, R.L.}, \bibinfo{author}{Shamir, A.},
  \bibinfo{author}{Adleman, L.}, \bibinfo{year}{1978}b.
\newblock \bibinfo{title}{A method for obtaining digital signatures and
  public-key cryptosystems}.
\newblock \bibinfo{journal}{Communications of the ACM} \bibinfo{volume}{21},
  \bibinfo{pages}{120--126}.
\bibitem[{Robinson et~al.(2009)Robinson, Graux, Botterman and
  Valeri}]{202_robinson2009review}
\bibinfo{author}{Robinson, N.}, \bibinfo{author}{Graux, H.},
  \bibinfo{author}{Botterman, M.}, \bibinfo{author}{Valeri, L.},
  \bibinfo{year}{2009}.
\newblock \bibinfo{title}{Review of the european data protection directive}.
\newblock \bibinfo{journal}{Rand Europe} .
\bibitem[{Shang et~al.(2020)Shang, Tang, Chen and Liu}]{147_shang2020full}
\bibinfo{author}{Shang, T.}, \bibinfo{author}{Tang, Y.}, \bibinfo{author}{Chen,
  R.}, \bibinfo{author}{Liu, J.}, \bibinfo{year}{2020}.
\newblock \bibinfo{title}{Full quantum one-way function for quantum
  cryptography}.
\newblock \bibinfo{journal}{Quantum Engineering} \bibinfo{volume}{2},
  \bibinfo{pages}{e32}.
\bibitem[{Shirazi et~al.(2018)Shirazi, Simeonovski, Asghar, Backes and
  Diaz}]{101_shirazi2018survey}
\bibinfo{author}{Shirazi, F.}, \bibinfo{author}{Simeonovski, M.},
  \bibinfo{author}{Asghar, M.R.}, \bibinfo{author}{Backes, M.},
  \bibinfo{author}{Diaz, C.}, \bibinfo{year}{2018}.
\newblock \bibinfo{title}{A survey on routing in anonymous communication
  protocols}.
\newblock \bibinfo{journal}{ACM Computing Surveys (CSUR)} \bibinfo{volume}{51},
  \bibinfo{pages}{1--39}.
\bibitem[{Sweeney(2002)}]{103_sweeney2002k}
\bibinfo{author}{Sweeney, L.}, \bibinfo{year}{2002}.
\newblock \bibinfo{title}{k-anonymity: A model for protecting privacy}.
\newblock \bibinfo{journal}{International Journal of Uncertainty, Fuzziness and
  Knowledge-Based Systems} \bibinfo{volume}{10}, \bibinfo{pages}{557--570}.
\bibitem[{Thaler et~al.(2012)Thaler, Ullman and Vadhan}]{127_thaler2012faster}
\bibinfo{author}{Thaler, J.}, \bibinfo{author}{Ullman, J.},
  \bibinfo{author}{Vadhan, S.}, \bibinfo{year}{2012}.
\newblock \bibinfo{title}{Faster algorithms for privately releasing marginals},
  in: \bibinfo{booktitle}{International Colloquium on Automata, Languages, and
  Programming}, \bibinfo{organization}{Springer}. pp.
  \bibinfo{pages}{810--821}.
\bibitem[{Ullman(2016)}]{41_ullman2016answering}
\bibinfo{author}{Ullman, J.}, \bibinfo{year}{2016}.
\newblock \bibinfo{title}{Answering $n^{2+o(1)}$ counting queries with
  differential privacy is hard}.
\newblock \bibinfo{journal}{SIAM Journal on Computing} \bibinfo{volume}{45},
  \bibinfo{pages}{473--496}.
\bibitem[{Ullman and Vadhan(2020)}]{40_ullman2020pcps}
\bibinfo{author}{Ullman, J.}, \bibinfo{author}{Vadhan, S.},
  \bibinfo{year}{2020}.
\newblock \bibinfo{title}{Pcps and the hardness of generating synthetic data}.
\newblock \bibinfo{journal}{Journal of Cryptology} , \bibinfo{pages}{1--35}.
\bibitem[{Vadhan(2017)}]{42_vadhan2017complexity}
\bibinfo{author}{Vadhan, S.}, \bibinfo{year}{2017}.
\newblock \bibinfo{title}{The complexity of differential privacy}, in:
  \bibinfo{booktitle}{Tutorials on the Foundations of Cryptography}.
  \bibinfo{publisher}{Springer}, pp. \bibinfo{pages}{347--450}.
\bibitem[{Van~Dijk et~al.(2010)Van~Dijk, Gentry, Halevi and
  Vaikuntanathan}]{160_van2010fully}
\bibinfo{author}{Van~Dijk, M.}, \bibinfo{author}{Gentry, C.},
  \bibinfo{author}{Halevi, S.}, \bibinfo{author}{Vaikuntanathan, V.},
  \bibinfo{year}{2010}.
\newblock \bibinfo{title}{Fully homomorphic encryption over the integers}, in:
  \bibinfo{booktitle}{Annual International Conference on the Theory and
  Applications of Cryptographic Techniques}, \bibinfo{organization}{Springer}.
  pp. \bibinfo{pages}{24--43}.
\bibitem[{Victor et~al.(2020)Victor, Frenkel and
  Kershner}]{233_victor2020personal}
\bibinfo{author}{Victor, D.}, \bibinfo{author}{Frenkel, S.},
  \bibinfo{author}{Kershner, I.}, \bibinfo{year}{2020}.
\newblock \bibinfo{title}{Personal data of all 6.5 million israeli voters is
  exposed}.
\newblock \bibinfo{journal}{The New York Times. https://www. nytimes.
  com/2020/02/10/world/middleeast/israeli-voters-leak. html} .
\bibitem[{Wagh et~al.(2018)Wagh, Cuff and Mittal}]{197_wagh2018differentially}
\bibinfo{author}{Wagh, S.}, \bibinfo{author}{Cuff, P.},
  \bibinfo{author}{Mittal, P.}, \bibinfo{year}{2018}.
\newblock \bibinfo{title}{Differentially private oblivious ram}.
\newblock \bibinfo{journal}{Proceedings on Privacy Enhancing Technologies}
  \bibinfo{volume}{2018}, \bibinfo{pages}{64--84}.
\bibitem[{Wagh et~al.(2020)Wagh, He, Machanavajjhala and Mittal}]{2_wagh2020dp}
\bibinfo{author}{Wagh, S.}, \bibinfo{author}{He, X.},
  \bibinfo{author}{Machanavajjhala, A.}, \bibinfo{author}{Mittal, P.},
  \bibinfo{year}{2020}.
\newblock \bibinfo{title}{Dp-cryptography: Marrying differential privacy and
  cryptography in emerging applications}.
\newblock \bibinfo{journal}{arXiv preprint arXiv:2004.08887} .
\bibitem[{Wang et~al.(2019)Wang, Xiao, Yang, Zhao, Hui, Shin, Shin and
  Yu}]{107_wang2019collecting}
\bibinfo{author}{Wang, N.}, \bibinfo{author}{Xiao, X.}, \bibinfo{author}{Yang,
  Y.}, \bibinfo{author}{Zhao, J.}, \bibinfo{author}{Hui, S.C.},
  \bibinfo{author}{Shin, H.}, \bibinfo{author}{Shin, J.}, \bibinfo{author}{Yu,
  G.}, \bibinfo{year}{2019}.
\newblock \bibinfo{title}{Collecting and analyzing multidimensional data with
  local differential privacy}, in: \bibinfo{booktitle}{2019 IEEE 35th
  International Conference on Data Engineering (ICDE)},
  \bibinfo{organization}{IEEE}. pp. \bibinfo{pages}{638--649}.
\bibitem[{Warner(1965)}]{221_warner1965randomized}
\bibinfo{author}{Warner, S.L.}, \bibinfo{year}{1965}.
\newblock \bibinfo{title}{Randomized response: A survey technique for
  eliminating evasive answer bias}.
\newblock \bibinfo{journal}{Journal of the American Statistical Association}
  \bibinfo{volume}{60}, \bibinfo{pages}{63--69}.
\bibitem[{Wu(2020)}]{176}
\bibinfo{author}{Wu, C.}, \bibinfo{year}{2020}.
\newblock \bibinfo{title}{relationship among various distributions}.
\newblock \bibinfo{howpublished}{[EB/OL]}.
\newblock \bibinfo{note}{\url{https://www.zhihu.com/question/36214010}}.
\bibitem[{Wu et~al.(2016)Wu, He, Wu and Xia}]{191_wu2016inherit}
\bibinfo{author}{Wu, G.}, \bibinfo{author}{He, Y.}, \bibinfo{author}{Wu, J.},
  \bibinfo{author}{Xia, X.}, \bibinfo{year}{2016}.
\newblock \bibinfo{title}{Inherit differential privacy in distributed setting:
  Multiparty randomized function computation}, in: \bibinfo{booktitle}{2016
  IEEE Trustcom/BigDataSE/ISPA}, \bibinfo{organization}{IEEE}. pp.
  \bibinfo{pages}{921--928}.
\bibitem[{Yang et~al.(2020)Yang, Lyu, Zhao, Zhu and Lam}]{214_yang2020local}
\bibinfo{author}{Yang, M.}, \bibinfo{author}{Lyu, L.}, \bibinfo{author}{Zhao,
  J.}, \bibinfo{author}{Zhu, T.}, \bibinfo{author}{Lam, K.Y.},
  \bibinfo{year}{2020}.
\newblock \bibinfo{title}{Local differential privacy and its applications: A
  comprehensive survey}.
\newblock \bibinfo{journal}{arXiv preprint arXiv:2008.03686} .
\bibitem[{Yao(1982)}]{193_yao1982protocols}
\bibinfo{author}{Yao, A.C.}, \bibinfo{year}{1982}.
\newblock \bibinfo{title}{Protocols for secure computations}, in:
  \bibinfo{booktitle}{23rd annual symposium on foundations of computer science
  (sfcs 1982)}, \bibinfo{organization}{IEEE}. pp. \bibinfo{pages}{160--164}.
\bibitem[{Yao(1986)}]{194_yao1986generate}
\bibinfo{author}{Yao, A.C.C.}, \bibinfo{year}{1986}.
\newblock \bibinfo{title}{How to generate and exchange secrets}, in:
  \bibinfo{booktitle}{27th Annual Symposium on Foundations of Computer Science
  (sfcs 1986)}, \bibinfo{organization}{IEEE}. pp. \bibinfo{pages}{162--167}.
\bibitem[{Yu(2016)}]{219_yu2016big}
\bibinfo{author}{Yu, S.}, \bibinfo{year}{2016}.
\newblock \bibinfo{title}{Big privacy: Challenges and opportunities of privacy
  study in the age of big data}.
\newblock \bibinfo{journal}{IEEE access} \bibinfo{volume}{4},
  \bibinfo{pages}{2751--2763}.
\bibitem[{Zhandry(2020)}]{148_zhandry2020new}
\bibinfo{author}{Zhandry, M.}, \bibinfo{year}{2020}.
\newblock \bibinfo{title}{New techniques for traitor tracing: Size $n^{1/3}$
  and more from pairings}, in: \bibinfo{booktitle}{Annual International
  Cryptology Conference}, \bibinfo{organization}{Springer}. pp.
  \bibinfo{pages}{652--682}.
\bibitem[{Zhang et~al.(2012)Zhang, Zhang, Xiao, Yang and
  Winslett}]{187_zhang2012functional}
\bibinfo{author}{Zhang, J.}, \bibinfo{author}{Zhang, Z.},
  \bibinfo{author}{Xiao, X.}, \bibinfo{author}{Yang, Y.},
  \bibinfo{author}{Winslett, M.}, \bibinfo{year}{2012}.
\newblock \bibinfo{title}{Functional mechanism: regression analysis under
  differential privacy}.
\newblock \bibinfo{journal}{arXiv preprint arXiv:1208.0219} .
\bibitem[{Zhao et~al.(2019)Zhao, Zhang, Wan, Liu and Umer}]{215_zhao2019survey}
\bibinfo{author}{Zhao, P.}, \bibinfo{author}{Zhang, G.}, \bibinfo{author}{Wan,
  S.}, \bibinfo{author}{Liu, G.}, \bibinfo{author}{Umer, T.},
  \bibinfo{year}{2019}.
\newblock \bibinfo{title}{A survey of local differential privacy for securing
  internet of vehicles}.
\newblock \bibinfo{journal}{The Journal of Supercomputing} ,
  \bibinfo{pages}{1--22}.
\bibitem[{Zhu et~al.(2017)Zhu, Li, Zhou and Philip}]{218_zhu2017differentially}
\bibinfo{author}{Zhu, T.}, \bibinfo{author}{Li, G.}, \bibinfo{author}{Zhou,
  W.}, \bibinfo{author}{Philip, S.Y.}, \bibinfo{year}{2017}.
\newblock \bibinfo{title}{Differentially private data publishing and analysis:
  A survey}.
\newblock \bibinfo{journal}{IEEE Transactions on Knowledge and Data
  Engineering} \bibinfo{volume}{29}, \bibinfo{pages}{1619--1638}.

\end{thebibliography}

\end{document}